\documentclass[12pt]{article}
\usepackage{epsfig,amssymb}
\usepackage{amsmath}
\def\bea{\begin{eqnarray}}
\def\eea{\end{eqnarray}}
\def\be{\begin{equation}}
\def\ee{\end{equation}}
\def\mathe{{\mathfrak e}}

\numberwithin{equation}{section}

\begin{document}
\title{
{\baselineskip -.2in
\vbox{\small\hskip 4in \hbox{}}}
\vskip .4in 
Counting Wobbling Dual-Giants}
\author{Sujay K. Ashok$^{a,b}$ and Nemani V. Suryanarayana$^a$ \\
{}\\
{\small{\it $^a$Institute of Mathematical Sciences}}\\
{\small{\it C.I.T Campus, Taramani}}\\
{\small{\it Chennai 600113, India}} \\
\\
{\small{\it $^b$Perimeter Institute for Theoretical Physics}}\\
{\small{\it Waterloo, Ontario, N2L 2Y5, Canada}}\\
\\
{\small{E-mail: {\tt sashok, nemani@imsc.res.in}}} \\
}
\date{}
\maketitle

\begin{abstract}

We derive the BPS equations for D3-branes embedded in $AdS_5 \times S^5$ that preserve at least two supercharges. These are given in terms of conditions on the pullbacks of some space-time differential four-forms. Solutions to our equations are shown to describe all the known giant and dual-giant gravitons in $AdS_5 \times S^5$. We then argue that the configuration spaces of dual-giants can be mapped to non-compact hyperbolic versions of complex projective spaces, in contrast with the giants, whose configuration spaces have been mapped to complex projective spaces. We quantize the configuration space of the 1/8-BPS dual-giants with two angular momenta in $AdS_5$ and one angular momentum in $S^5$ and find agreement with the partition function in the literature obtained both from counting appropriate 1/8-BPS configurations of giants and the boundary gauge theory considerations. 
\end{abstract}

\newpage

\tableofcontents

\section{Introduction}

The program of counting supersymmetric states in the context of AdS/CFT correspondence is an important one as it helps in verifying the AdS/CFT correspondence in its BPS sector. In recent times a lot of progress has been made, both from the bulk and the boundary point of view, in cases where the states under consideration preserve at least four supersymmetries.

In the well studied maximally supersymmetric version of the AdS/CFT correspondence \cite{maldacena}, the bulk theory is the type IIB string theory on  $AdS_5 \times S^5$ background and the corresponding boundary theory is the 4-dimensional ${\cal N} = 4$ $U(N)$ SYM on $S^3 \times \mathbb{R}$. In this case any state on the string theory side can be specified by six conserved charges $(E, S_1, S_2, J_1, J_2, J_3)$ where ($E, S_1, S_2$) denote the energy and the two angular momenta in global $AdS_5$, and ($J_1, J_2, J_3$) denote the three independent angular momenta on $S^5$, namely, the R-charges. The BPS states satisfy a linear relation among these six charges. On the bulk side some of the bosonic BPS states with non-zero charges $(J_1, J_2, J_3)$ at the classical level are given by finite-energy D3-brane configurations in $AdS_5 \times S^5$ \cite{mst, gmt, hhi, mikhailov, rcmnvs, gmnvs, kimlee2}. They turn out to have two independent and equivalent description in terms of the Mikhailov giant gravitons \cite{mikhailov} and the dual-giant graviton configurations of \cite{gmnvs}. Their quantization was carried out in \cite{beasley, bglm, gmnvs} and shown to reproduce the partition function of chiral primaries \cite{kmmr} obtained from the bosonic fields on the CFT side. For earlier work leading to the quantization of giants and dual-giants in $AdS_5 \times S^5$, see \cite{djm, gautam, nvs, adkss} and see \cite{sbsm, mrs} for giants in other contexts. 

However, such states are not useful in making progress with the problem of accounting for the entropy of five dimensional supersymmetric black holes of \cite{gr1, gr2}. These black hole solutions can be lifted to black holes in $AdS_5 \times S^5 $ background of the type IIB string theory. They preserve only two supersymmetries \cite{ggs} and necessarily carry non-zero angular momenta ($S_1, S_2$). The microstates of these black holes will be the 1/16-BPS states in type IIB string theory on $AdS_5 \times S^5$ or equivalently the 1/16-BPS states of $d=4$, ${\cal N} = 4$, $U(N)$ SYM on $S^3 \times \mathbb{R}$. Recently some important progress has been reported in \cite{ggkm} where the partition function of 1/16-BPS operators made out of the bosonic fields in ${\cal N} = 4$ $U(N)$ SYM, has been written down in some cases. These CFT states should be some of the microstates of the 1/16-BPS black holes. It is of interest to ask if the results of \cite{ggkm} can be reproduced by counting an appropriate set of 1/16-BPS states in type IIB theory on the $AdS_5 \times S^5$ background with all possible charges ($S_1, S_2, J_1, J_2, J_3$). Of course there already exists a nice description of D3-brane configurations of $AdS_5 \times S^5$ with at least two supersymmetries by Kim and Lee \cite{kimlee2}. Their quantization, however, remains an open problem (see \cite{ggkm} for some approxmimate results).

So far, the only BPS D3-branes with non-zero $S_1$ and $S_2$ whose quantization has been achieved are described in terms of the 1/8-BPS configurations of 1/2-BPS giant gravitons \cite{gmnvs}. Their partition function turns out to be the same as that of the configurations of arbitrary number of bosons in a 3-dimensional harmonic oscillator with the restriction that the level number of one of the oscillators be less than or equal to $N$. It is interesting to ask if the duality between giants and the dual-giants that seems to exist (see for instance \cite{dms}) in the non-zero ($J_1, J_2, J_3$) sector also exists in the ($S_1, S_2, J_1$) sector or not. For this one would like to know if one can recover the partition function of the ($S_1, S_2, J_1$) states from some dual-giant like objects. In this paper we show that the wobbling dual-giants of \cite{rcmnvs} with quantum numbers $(S_1, S_2, J_1)$ give rise to the same partition function as the one computed in \cite{gmnvs} from the giant perspective. 

The derivations of supersymmetric giants in \cite{mikhailov, rcmnvs, kimlee2} start with spinors in an auxiliary 12-dimensional Lorentian geometry with two time-like directions: 
\be
\mathbb{C}^{1,2} \times \mathbb{C}^3\quad \text{with coordinates}\quad (\Phi_0, \Phi_1, \Phi_2; Z_1, Z_2, Z_3)\,,
\ee 
where $AdS_5 \times S^5$ is given by two real conditions 
\be
-|\Phi_0|^2 + |\Phi_1|^2 + |\Phi_2|^2 = - l^2 \quad \text{and}\quad |Z_1|^2 + |Z_2|^2 + |Z_3|^2 = l^2\,.
\ee 
Though a very effective method for $AdS_5 \times S^5$, it does not extend to cases such as the $\beta$-deformed backgrounds where such an embedding into a more tractable ambient space is not readily available. 

So it is desirable to develop techniques which will enable one to obtain the supersymmetric solutions using more direct means. With this motivation, in the first part of this paper, we reexamine the 1/16-BPS probe D3-branes in $AdS_5 \times S^5$. We start directly with the Killing spinor of $AdS_5 \times S^5$ and impose a set of projection conditions which reduces the number of independent components of the  Killing spinor to just two. Using the $\kappa$--projection condition on the world-volume of a generic D3-brane embedding, we derive the full set of BPS equations for the embedding coordinates. These are given in terms of a set of vanishing conditions on some four-forms in $AdS_5 \times S^5$ geometry. As we shall show, our procedure also makes manifest the calibrating forms for giant and dual-giant gravitons \cite{h-js}.  We illustrate that the Mikhailov giants \cite{mikhailov} and the wobbling dual-giants of \cite{kimlee2, rcmnvs} are solutions to our set of BPS equations and compare our analysis with the results of Kim and Lee \cite{kimlee2}.

In the second part of this paper we analyze particular classes of wobbling dual-giants carrying non-zero $(S_1, S_2, J_1)$ charges. We give a prescription to implement the stringy exclusion principle generalizing the one for the 1/2-BPS dual-giants \cite{gmnvs} to the $(S_1, S_2, J_1)$ dual-giant case. To quantize such classes of classical solutions one first computes the symplectic structure on their parameter space treated as a classical phase space. The symplectic form on this phase space can be derived using the covariant phase space methods of Crnkovic, Witten and Zuckerman \cite{Crnkovic, Zuckerman}. This technique has been used successfully in \cite{bglm} to compute the symplectic structure on the configuration space of the Mikhailov giants which turned out to be the complex projective space $\mathbb{CP}^n$ where $n$ is a regulator. Here we suitably adapt their techniques to the $(S_1, S_2, J_1)$ dual-giants. 

We show that the moduli-space of these wobbling dual-giants is generically non-compact. We argue that the configuration space of these objects as a phase space can be mapped to the hyperbolic version of the complex projective space, which we denote by $\widetilde{\mathbb{CP}}^n$ where $n$ is again a regulator. The symplectic structure turns out to be the the kahler form on these non-compact Kahler manifolds. We quantize these spaces using the methods of (holomorphic) geometric quantization. After removing the regulator, we are able to recover the partition function of the $(S_1, S_2, J_1)$ giants obtained earlier in \cite{gmnvs}, thereby giving evidence for the existence of a duality between the giants and dual-giants in this 1/8-BPS sector. 

\vskip .5cm
\noindent{\bf Organization of the paper}: In Section \ref{killspinor}, we use the explicit form of the Killing spinor in $AdS_5 \times S^5$ and analyze the kappa symmetry conditions for embedding supersymmetric D3-branes in the background. We obtain the constraints on the pull-back of space-time $4$-forms that lead to classical $1/16$-supersymmetric solutions. We derive the well known classical solutions that corresspond to the case of giants and dual-giants in a covariant manner. We then study dual-giants in Section \ref{charges} and derive the expressions for the charges that follow from the combined DBI and Wess-Zumino terms in the action for a probe D3-brane. Our discussion is very general in this section and not dependent on specific solutions to the equations of motion. In section \ref{symplectic}, we identify the classical solution space with the phase space of the classical theory and compute the symplectic structure on this phase space for a few chosen examples. This will prove crucial in Section \ref{quantum}, when we turn to the geometric quantization of the phase space and obtain the partition function of the $(S_1,S_2,J_1)$ dual-giants. In Section \ref{discuss}, we discuss possible future applications of our present work and mention some relations of our work to the existing literature. 
Some technical details pertaining to the computation of charges are collected in Appendix A. We collect some basic facts regarding the hyperbolic space $\widetilde{\mathbb{CP}}^m$ and discuss holomprhic quantization of this K\"ahler manifold in Appendix B.  

\section{Classical description of $\frac{1}{16}$-BPS giants}\label{killspinor}

In this section we derive the BPS equations for a general configuration of a D3-brane preserving at least 2 of the supersymmetries of $AdS_5 \times S^5$ by analyzing the $\kappa$-projection conditions for the D3-brane world-volume theory. Then we solve them restricting our attention to giant-like and dual-giant-like configurations.

\subsection{The 1/16-BPS equations}

We begin by studying the kappa-projection conditions that ensure supersymmetry for a D3-brane embedded in $AdS_5\times S^5$. For this we take the metric on $AdS_5 \times S^5$ written in global coordinates to be 
\begin{multline}
\frac{ds^2}{l^2} = -(1+\frac{r^2}{l^2})d\phi_0^2 + \frac{dr^2}{r^2+l^2}+\frac{r^2}{l^2}(d\theta^2+\cos^2\theta d\phi_1^2+\sin^2\theta d\phi_2^2)\cr
+ d\alpha^2 + \sin^2\alpha d\xi_1^2+\cos^2\alpha(d\beta^2 +\sin^2\beta d\xi_2^2+\cos^2\beta d\xi_3^2)\,,
\end{multline}
where $\phi_0 = \frac{t}{l}$. We choose the following frame for the $AdS_5$ part of the metric
\begin{align}\label{adsframe}
e^0 &= l \, V \, d \phi_0 - \textstyle{\frac{r^2}{l}} (\cos^2\theta d\phi_1 + \sin^2\theta d \phi_2), \cr 
e^1 &= V^{-1/2} \, dr\,,\qquad e^2 = r \, d\theta, \cr
e^3 &= r \, V^{1/2} ( \cos^2\theta ~ d\phi_{01} + \sin^2\theta  ~ d\phi_{02})\, \cr
e^4 &= r \, \cos\theta \sin\theta ~ d\phi_{12}\,,
\end{align}
where $V = 1+ r^2/l^2$ and $\phi_{ij} = \phi_i - \phi_j$. Here the ranges of various coordinates are: $-\infty < \phi_0 < \infty$, $0\le r < \infty$, $0 \le \theta \le \pi/2$, $0 \le \phi_1, \phi_2 < 2\pi$. This frame makes manifest the fact that $AdS_5$  can be written as a U(1) Hopf fibration: the base being the hyperbolic K\"ahler manifold $\widetilde{\mathbb{CP}}^2$, spanned by the  $\{r, \theta, \phi_{01}, \phi_{02}\}$ coordinates, and the fibre along the $\phi_0+\phi_1+\phi_2$ direction. For the $S^5$ part, we choose the frame
\begin{align}\label{sframe}
e^5 &= l \, d\alpha, \qquad e^6 = l \, \cos\alpha \, d\beta, \cr
e^7 &= l \, \cos\alpha \sin\alpha \, (\sin^2\beta \, d\xi_{12} + \cos^2\beta \, d\xi_{13}), \cr
e^8 &= l \, \cos\alpha \cos\beta \sin\beta \, d\xi_{23}, \cr
e^9 &= l \, (\sin^2\alpha \, d\xi_1 + \cos^2\alpha \sin^2\beta \, d\xi_2 +  \cos^2\alpha \cos^2\beta \,d\xi_3).
\end{align}
where $\xi_{ij} = \xi_i - \xi_j$ and the ranges of the coordinates are: $0 \le \alpha, \beta \le \pi/2$, $0 \le \xi_i < 2\pi$.  Again the choice of frame exhibits the fact that $S^5$ is a Hopf fibration over the K\"ahler manifold $\mathbb{CP}^2$ spanned by the $\{\alpha, \beta, \xi_{12}, \xi_{13}\}$ coordinates, with the fibre direction along $\xi_1+\xi_2+\xi_3$. 

The Killing spinor for the $AdS_5 \times S^5$ background adapted to the above frame  is given by   
\begin{multline}
\label{adskss2}
\epsilon = ~ e^{-\frac{1}{2} (\Gamma_{79} - i \Gamma_5 \, \tilde \gamma) \, \alpha}
e^{-\frac{1}{2} (\Gamma_{89} - i \Gamma_6 \tilde \gamma) \beta} \, e^{\frac{1}{2} 
\xi_1 \Gamma_{57}} \, e^{\frac{1}{2} \xi_2  \Gamma_{68}} \,
e^{\frac{i}{2} \xi_3  \Gamma_9 \, \tilde \gamma} \cr
\times e^{\frac{1}{2} \sinh^{-1}\! (\frac{r}{l}) \, (\Gamma_{03} + i \Gamma_1 \,
\gamma)} \,  e^{\frac{1}{2}
\theta \, (\Gamma_{12} + \Gamma_{34})} \, e^{\frac{i}{2} \phi_0 \, \Gamma_0 \, \gamma} \, e^{-\frac{1}{2} \phi_1 \Gamma_{13}} \,
e^{-\frac{1}{2} \phi_2 \Gamma_{24}} \, \epsilon_0
\end{multline}
where $\epsilon_0$ is an arbitrary 32 component weyl spinor satisfying $\Gamma_0 \cdots \Gamma_9 \epsilon_0 = - \epsilon_0$ and $\gamma = \Gamma^{01234}$, $\tilde\gamma = \Gamma^{56789}$ as in \cite{gmnvs}. We seek the full set of BPS equations for D3-branes in $AdS_5 \times S^5$ which preserve two supersymmetries out of the full  set of thirty two. Clearly this choice is non-unique. Without loss of generality we could choose them to be the ones obtained in \cite{gmnvs}. So we take superymmetries preserved by the D3-brane to be those that survive the projections
\begin{equation}
\label{projections}
\Gamma_{57} \epsilon_0 = \Gamma_{68} \epsilon_0 = i \epsilon_0\,, \qquad \Gamma_{09} \epsilon_0 =  -\epsilon_0\,,\qquad \Gamma_{13} \epsilon_0 = \Gamma_{24} \epsilon_0 = -i \epsilon_0. 
\end{equation}
With these projections the killing spinor simplifies to
\begin{equation}
\label{reducedks}
\epsilon = e^{\frac{i}{2} (\phi_0 + \phi_1 + \phi_2 + \xi_1 + \xi_2 + \xi_3) } \epsilon_0.
\end{equation}
Next we seek the equations that any D3-brane should satisfy to preserve (at least) these two supersymmetries. The ansatz we take for the D3-brane is the most general one, such that all the coordinates $(t, r, \theta, \phi_1, \phi_2, \alpha, \beta, \xi_1, \xi_2, \xi_3)$ are functions of the world-volume coordinates $(\tau, \sigma_1, \sigma_2, \sigma_3)$. The world-volume gamma matrices are
\begin{equation}
\gamma_i = {\mathfrak e}^a_i\, \Gamma_a
\end{equation}
where ${\mathfrak e}^a_i = e^a_\mu\, \partial_i X^\mu$, with $i \in \{\tau, \sigma_1, \sigma_2, \sigma_3\}$, is the pull-back of $e^a_\mu$ onto the world-volume. Then we have 
\begin{equation}
\gamma_{\tau \sigma_1 \sigma_2 \sigma_3} = {\mathfrak e}_0^a\, {\mathfrak e}_1^b\, {\mathfrak e}_2^c\,  {\mathfrak e}_3^d\, \Gamma_{abcd}.
\end{equation}
The kappa projection condition on the worldvolume of the D3-brane is given by
\begin{equation}\label{kappaeqn}
\gamma_{\tau \sigma_1 \sigma_2 \sigma_3} \epsilon = \pm i \, \sqrt{-\det h} ~ \epsilon \,,
\end{equation}
where the sign distinguishes a D3-brane from an anti-D3-brane. To obtain the equations that the $\kappa$-projection condition implies we substitute the spinor in \eqref{reducedks} into \eqref{kappaeqn} and simplify the left hand side of \eqref{kappaeqn} using the projections \eqref{projections} untill it reduces to a linear combination of the independent column matrices of the type $\Gamma_{ab \cdots} \epsilon_0$. Such column matrices naturally fall into two types. Ones in which at least one $\Gamma$-matrix multiplies $\epsilon_0$ and other where no $\Gamma$-matrix (but only the identity matrix) multiplies $\epsilon_0$. Then we simply have to set the coefficient of each such independent column matrix to zero.\footnote{This technique was first used in \cite{gmnvs}, albeit in a much simpler context.} 

In order to write down the BPS equations in a compact form, let us define the complex one-forms
\begin{equation}\label{cplxforms}
{\bf E}^1 = {\mathfrak e}^1-i {\mathfrak e}^3\qquad {\bf E}^2 = {\mathfrak e}^2-i {\mathfrak e}^4\qquad {\bf E}^5= {\mathfrak e}^5+i {\mathfrak e}^7\qquad {\bf E}^6= {\mathfrak e}^6+i {\mathfrak e}^8 \,,
\end{equation}
along with the real 1-forms 
\begin{equation}\label{realforms}
{\bf E^0} = {\mathfrak e}^0 + {\mathfrak e}^9 \qquad\text{and}\qquad {\bf E}^{\bar 0} = \mathe^0 - \mathe^9 \,
\end{equation}
as well as the two-forms
\begin{align}
\tilde{\bf \omega} &= {\mathfrak e}^{13}+ {\mathfrak e}^{24}= -\frac{i}{2}\left({\bf E}^1\wedge \overline{{\bf E}^1} + {\bf E}^2\wedge \overline{{\bf E}^2}\right) \equiv \omega_{\widetilde{\mathbb{CP}}^2}\cr
{\bf \omega}&= {\mathfrak e}^{57}+ {\mathfrak e}^{68}=\frac{i}{2}\left({\bf E}^5\wedge \overline{{\bf E}^5} + {\bf E}^6\wedge \overline{{\bf E}^6}\right) \equiv \omega_{\mathbb{CP}^2}\,.
\end{align}
The $2$-forms are the pull-backs onto the worldvolume of the brane, of the K\"ahler forms on the respective base manifolds $\mathbb{CP}^2$ and $\widetilde{\mathbb{CP}}^2$ when $S^5$ and $AdS_5$ are written as Hopf-fibrations. With these definitions, the BPS equations that follow from \eqref{kappaeqn} by setting the coefficient of a column matrix of the type $\Gamma_{ab\cdots} \epsilon_0$ with at least one $\Gamma$-matrix can be written in the following compact form:
\begin{align}
\label{compactbps}
{\bf E}^{1256}&= 0 \cr
({\mathfrak e}^0+{\mathfrak e}^9)\wedge {\bf E}^{ABC} &= 0 \quad \text{and}\cr 
({\mathfrak e}^{09}+i\, (\tilde{\bf \omega}- {\bf \omega})) \wedge {\bf E}^{AB} &= 0 \quad \hbox{for} \quad A, B = 0, 1, 2, 5, 6\, .
\end{align}
Substituting these equations into the kappa projection equation \eqref{kappaeqn}, we get the equation
\begin{equation}\label{identitycondition}
{\mathfrak e}^{09}\wedge (\tilde{\bf \omega} - {\bf \omega}) + \frac{i}{2} (\tilde{\bf \omega}-{\bf \omega}) \wedge (\tilde{\bf \omega} -{\bf \omega}) = \pm \sqrt{-\det h} 
\end{equation}
as the coefficient of $\epsilon_0$ (with no $\Gamma$-matrix multiplying it). We still need to check if the equations \eqref{compactbps} are sufficient to satisfy this equation identically for either sign on the right hand side. To simplify the right hand side, we first note the identity 
\begin{align}
-\det h &= -\det_{ij}\left(\sum_{m,n=0}^{9}{\mathfrak e}^m_{i}{\mathfrak e}^n_{j}\, \eta_{mn}\right) \cr
&= -\sum_{m_i<n_i<p_i<q_i}(\epsilon^{i_1j_1k_1l_1}\, {\mathfrak e}^{m_1}_{i_1}{\mathfrak e}^{n_1}_{j_1}{\mathfrak e}^{p_1}_{k_1}{\mathfrak e}^{q_1}_{k_1})\, (\epsilon^{i_1j_2k_2l_2}\, {\mathfrak e}^{m_2}_{i_2}{\mathfrak e}^{n_2}_{j_2}{\mathfrak e}^{p_2}_{k_2} {\mathfrak e}^{q_2}_{l_2}) 
\eta_{m_1m_2}\eta_{n_1n_2}\eta_{p_1p_2}\eta_{q_1q_2}\cr
& = \sum_{a<b} {\mathfrak e}^{09ab} {\mathfrak e}^{09ab}+ \sum_{a<b<c} ({\mathfrak e}^{0abc} {\mathfrak e}^{0abc}- {\mathfrak e}^{9abc} {\mathfrak e}^{9abc})-
\sum_{a<b<c<d} {\mathfrak e}^{abcd} {\mathfrak e}^{abcd} \,.
\end{align}
Using the useful identity
\begin{equation}
{\mathfrak e}^{abc[d} {\mathfrak e}^{ef]ab} = 0 \,,
\end{equation}
where $a,b$ need not be summed,  one can rewrite each of the terms in the determinant in terms of the pullback of the complex one-forms $E^A$ as follows:
\begin{align}
\sum_{a<b} {\mathfrak e}^{09ab} {\mathfrak e}^{09ab} &= \sum_{A<B}\vert {\mathfrak e}^{09}\wedge {\bf E}^{AB} \vert^2 + \vert {\mathfrak e}^{09}\wedge(\tilde{\bf \omega} -{\bf \omega}) \vert^2 \cr
\sum_{a<b<c} {\mathfrak e}^{*abc} {\mathfrak e}^{*abc} &= \sum_{A<B<C} \vert {\mathfrak e}^* \wedge {\bf E}^{ABC}\vert^2 + \sum_{A} \vert {\mathfrak e}^* \wedge {\bf E}^A \wedge (\tilde{\bf \omega} - {\bf \omega})\vert^2\,,  \quad \text{where}\quad \text{*}\in \{0,9\} \cr
\sum_{a<b<c<d} {\mathfrak e}^{abcd} {\mathfrak e}^{abcd} &= \vert {\bf E}^{1256}\vert^2 + \sum_{A<B}\vert {\bf E}^{AB} \wedge (\tilde{\bf \omega}-{\bf \omega})\vert^2+\frac{1}{4}\vert (\tilde{\bf \omega}-{\bf \omega})\wedge (\tilde{\bf \omega}-{\bf \omega})\vert^2 \,.
\end{align}
Now, using the BPS conditions \eqref{compactbps}, one can check that the determinant reduces to 
\begin{equation}
-\det h = \big( {\mathfrak e}^{09}\wedge (\tilde{\bf \omega}-{\bf \omega}) \big)^2 -\frac{1}{4}\left( (\tilde{\bf \omega}-{\bf \omega})\wedge (\tilde{\bf \omega}-{\bf \omega}) \right)^2 \,.
\end{equation}
Substituting this expression into \eqref{identitycondition} we see that it can not be satisfied without supplying further conditions. There are two ways we can solve the equation \eqref{identitycondition} which we refer to as ``time-like" and ``instantonic". 

For ``time-like" D3-branes we further impose the condition
\begin{equation}
\label{timelikebps}
({\bf \omega} - \tilde{\bf \omega})\wedge ({\bf \omega}-\tilde{\bf \omega}) = 0\,.
\end{equation}
Then $-(\det h)$ reduces to a complete square and from \eqref{identitycondition} we have 
\begin{equation}
{\mathfrak e}^{09} \wedge (\tilde{\bf \omega} - {\bf \omega}) = \pm \left\vert {\mathfrak e}^{09} \wedge (\tilde{\bf \omega}-{\bf \omega}) \right\vert = \pm \text{dvol}_4.
\end{equation}
This is solved for either branes or anti-branes depending on the sign of  $\vert {\mathfrak e}^{09} \wedge (\tilde{\bf \omega} -{\bf \omega})\vert$. Note that we have identified the $4$-volume element on the world volume of the D3-brane. We will point out its relation to the calibration forms on giant gravitons and dual-giant gravitons in the discussion section. 

It is amusing to note that our analysis suggests another class of supersymmetric ``instantonic'' branes if we choose 
\begin{equation}
{\mathfrak e}^{09} \wedge (\tilde{\bf \omega}-{\bf \omega}) = 0 \,.
\end{equation}
In this case the world-volume is given by the pull-back of $\frac{i}{2}(\tilde\omega - \omega) \wedge (\tilde\omega - \omega)$ which does not have a component along the time-like 1-form $e^0$. As we will see shortly, all the known solutions of giants and dual-giants are in the ``time-like" case. Therefore we shall restrict our analysis to the time-like D3-branes and analyze the BPS equations \eqref{compactbps} along with \eqref{timelikebps} for solutions.

One can immediately find two classes of solutions to the BPS equations that are usually referred to as either giants or dual-giants. Giants are those configurations that are point-like in the $\{0,1,2,3,4\}$ directions, so that the pullback of a form with more than one index from this set onto the world-volume vanishes. With this restriction, the list of BPS conditions for giants simplifies to 
\begin{align}
\label{fullggeqns}
{\mathfrak e}^{09} \wedge {\bf E}^{56} &= 0 \cr
({\mathfrak e}^0 + {\mathfrak e}^9) \wedge {\bf E}^{56} \wedge \begin{Bmatrix} {\bf E}^{1} \\ {\bf E}^{2} \end{Bmatrix} &= 0 \cr 
\begin{Bmatrix} {\mathfrak e}^0 + {\mathfrak e}^9 \\ {\bf E}^5 \\ {\bf E}^6 \end{Bmatrix} \wedge \begin{Bmatrix}  {\mathfrak e}^0 + {\mathfrak e}^9 \\  {\bf E}^1 \\ {\bf E}^2 \end{Bmatrix} \wedge {\bf \omega} &= 0\cr
{\bf \omega}\wedge {\bf \omega} = {\mathfrak e}^{5678} &= 0  \,.
\end{align}
Similarly for dual-giants, whose world-volume is point-like along the $\{5,6,7,8,9\}$ directions, the BPS conditions are
\begin{align}
\label{fulldggeqns}
{\mathfrak e}^{09} \wedge {\bf E}^{12} &= 0 \cr
({\mathfrak e}^0 + {\mathfrak e}^9) \wedge {\bf E}^{12} \wedge \begin{Bmatrix} {\bf E}^{5} \\ {\bf E}^{6} \end{Bmatrix} &= 0 \cr 
\begin{Bmatrix}  {\mathfrak e}^0 + {\mathfrak e}^9 \\  {\bf E}^5 \\ {\bf E}^6 \end{Bmatrix} \wedge \begin{Bmatrix} {\mathfrak e}^0 + {\mathfrak e}^9 \\ {\bf E}^1 \\ {\bf E}^2 \end{Bmatrix} \wedge \tilde {\bf \omega}  &= 0\cr
\tilde{\bf \omega} \wedge \tilde{\bf \omega} = {\mathfrak e}^{1234} &= 0  \,.
\end{align}

We will present explicit solutions for these conditions in the following sections. There are four classes of 1/8-BPS solutions: these can be classified by their quantum numbers and whether they satisfy \eqref{fullggeqns} or \eqref{fulldggeqns}. Let us mention these briefly before we proceed.

\begin{itemize}
\item $(J_1, J_2, J_3)$ dual-giants satisfy \eqref{fulldggeqns}, with spins only along the $S^5$. These were quantized in \cite{gmnvs} and it was shown that their partition function matches exactly with the partition function of the Mikhailov giants obtained in \cite{bglm}. 

\item $(J_1, J_2, J_3)$ giants satisfy \eqref{fullggeqns} and have spins only along the $S^5$ directions. These are the familiar Mikhailov giants \cite{mikhailov}. The quantization of their configuration space has been carried out in \cite{bglm}.

\item $(S_1, S_2, J_1)$ dual-giants (or wobbling dual-giants \cite{rcmnvs}), with two spins in $AdS_5$ and one spin along the $S^5$ will be the main focus of our paper. The configuration space of these dual-giants and its quantization is an open problem and will be addressed in the following sections. As expected, the partition function obtained coincides with the partition function obtained for the $(S_1, S_2, J_1)$ giants in \cite{gmnvs}. 

\item $(S_1, S_2, J_1)$ giants (or spinning giants \cite{adkss, cs, gmnvs})  satisfy \eqref{fullggeqns} and have two spins in the $AdS_5$ directions $(\phi_1, \phi_2)$ and one spin along the $\xi_1$ direction in the $S^5$. Their quantization has been carried out in \cite{gmnvs}.

\end{itemize}

Apart from these 1/8-BPS states, we will also discuss generalizations of these to 1/16-BPS configurations that describe a single giant or a single dual-giant. We will comment briefly on their quantization problem in the discussion section. 

\subsection{Dual-giant solutions}

The dual-giant is a generic term to describe D3-brane solutions which are point-like in the $S^5$ at any instant of the world-volume time. As a preliminary check on our BPS equations, let us affirm that the spherical dual-giant gravitons that have been discussed in \cite{gmnvs} are solutions to our BPS equations \eqref{fulldggeqns}. This will be useful to our more general discussion to follow. 

\subsubsection{$(J_1, J_2, J_3)$ dual-giant}\label{rounds3}

We use the ansatz $\phi_0 = \tau, \, \theta = \sigma_1, \, \phi_1 = \sigma_2, \, \phi_2 = \sigma_3$. Using these, one can write the pullbacks of the various vielbeins as follows:
\begin{align}
{\mathfrak e}^0 &= l\, V d\tau-\frac{r^2}{l}(\cos^2\sigma_1 d\sigma_2 + \sin^2\sigma_1 d\sigma_3)\qquad 
{\mathfrak e}^1 = \frac{\dot{r}d\tau}{V^{\frac{1}{2}}}, ~~ {\mathfrak e}^2 = r\, d\sigma_1, \cr
{\mathfrak e}^3 &= r V^{\frac{1}{2}}(d\tau-\cos^2\sigma_1 d\sigma_2-\sin^2\sigma_1 d\sigma_3), ~~
{\mathfrak e}^4 = r\cos\sigma_1\sin\sigma_1(d\sigma_2-d\sigma_3), \cr
{\mathfrak e}^5 &= l\, \dot{\alpha}\, d\tau, ~~
{\mathfrak e}^6 = l\, \cos\alpha\, \dot{\beta}\, d\tau, ~~
{\mathfrak e}^7 = l\cos\alpha\sin\alpha(\dot{\xi_1}-\sin^2\beta\, \dot{\xi_2}-\cos^2\beta\, \dot{\xi_3})d\tau, \cr
{\mathfrak e}^8 &= l\cos\alpha\cos\beta\sin\beta(\dot{\xi_2}-\dot{\xi_3})d\tau, ~~
{\mathfrak e}^9 = l (\sin^2\alpha\, \dot{\xi_1}+\cos^2\alpha\sin^2\beta\, \dot{\xi_2}+\cos^2\alpha\cos^2\beta\, \dot{\xi_3})d\tau \,. \nonumber
\end{align}
Consider the last equation ${\mathfrak e}^{1234}=0$ in \eqref{fulldggeqns} we find:
\begin{equation}
r^3\, \dot{r}\, \sin\sigma_1\cos\sigma_1\, d\tau\wedge d\sigma_1\wedge d\sigma_2\wedge d\sigma_3 = 0 \,. 
\end{equation}
This is solved only for $\dot{r}=0$. Thus, we find that 
\begin{equation}
{\mathfrak e}^1 = 0 \,.
\end{equation}
The first two equations in \eqref{fulldggeqns} are automatically satisfied using our ansatz. The third equation, $({\mathfrak e}^0+{\mathfrak e}^9)\wedge {\bf E}^1\wedge {\bf \omega} = 0$, using $\dot{r}=0$, reduces to the equation
\begin{equation}
1+\sin^2\alpha\, \dot{\xi_1}+\cos^2\alpha\sin^2\beta\, \dot{\xi_2}+\cos^2\alpha\cos^2\beta\, \dot{\xi_3} = 0 \,.
\end{equation}
Since we expect the solution to exist for arbitrary values of $\alpha$ and $\beta$, the solution is given by
\begin{equation}\label{xidots}
\dot{\xi_i} = -1 \quad \forall \quad i \in \{1,2,3\} \,.
\end{equation}
The remaining non-trivial equations lead to the conditions (here, we have used \eqref{xidots})
\begin{equation}
\begin{Bmatrix} \dot{\alpha}\, d\tau \\ \cos\alpha\,  \dot{\beta}\, d\tau \end{Bmatrix}  \wedge (r^3\, V^{\frac{1}{2}}\, \sin\sigma_1\cos\sigma_1)\, d\sigma \wedge d\sigma_2 \wedge d\sigma_3 = 0 \,,
\end{equation}
Thus, the full set of conditions that follow from our BPS equations is
\begin{equation}
\label{mseqns}
\dot{r}=\dot{\alpha}=\dot{\beta}=0 \qquad \text{and}\qquad \dot{\xi_1}=\dot{\xi_2}=\dot{\xi_{3}} = -1 \,. 
\end{equation}
We have thus recovered the conditions for supersymmetry derived in \cite{gmnvs} using our BPS equations \eqref{fulldggeqns}. The solution to these equations is 
\begin{equation}
\label{mssoln1}
r = r^{(0)}, ~~ \alpha = \alpha^{(0)}, ~~ \beta = \beta^{(0)}, ~~ \xi_i = \xi_i^{(0)} - \tau
\end{equation}
with six constant parameters $r^{(0)}, \, \alpha^{(0)}, \, \beta^{(0)}, \, \xi_i^{(0)}$ as in \cite{gmnvs}. This provides a good first consistency check of our equations. 

\subsubsection{$(S_1, S_2, J_1)$ ``wobbling'' dual-giants} 

We now try to generalize the solutions in the previous subsection and solve the list of BPS conditions for particular classes of dual-giants satisfying \eqref{fulldggeqns}. The general solution of the BPS conditions will be given by three complex conditions that will lead to a four dimensional world-volume \cite{kimlee2}. Let us consider one of the constraints and let it be a completely general function of all the coordinates
\begin{equation}
F(r,\theta, \alpha, \beta, \phi_0, \phi_1, \phi_2, \xi_1, \xi_2, \xi_3) = 0 \,.
\end{equation}
This leads to the differential constraint
\begin{align}
{\rm P}\left[F_r\, dr + F_{\theta}\, d\theta + F_{\alpha}\, d\alpha + F_{\beta}\, d\beta + \sum_{i=0,1,2} F_{\phi_i}\, d\phi_i + \sum_{i=1,2,3}F_{\xi_i}\, d\xi_i\right] = 0 
\end{align}
where ${\rm P}$ denotes pullback onto the world-volume. It is possible to rewrite each of these one forms in terms of the complex one-forms \eqref{cplxforms} using the explicit frames used in \eqref{adsframe} and \eqref{sframe}. This leads to the differential constraint 
\begin{multline}\label{dFeqn}
\left[ F_{\rho} -  i\, \big(\tanh\rho\, F_{\phi_0} + \coth\rho\, (F_{\phi_1}+F_{\phi_2})\big) \right]\, {\bf E}^1 + 
\left[ F_{\rho} + i\, \big(\tanh\rho\, F_{\phi_0} + \coth\rho\, (F_{\phi_1}+F_{\phi_2})\big) \right]\, \overline{\bf E^1}\cr
+ \frac{1}{\sinh\rho}\left[ F_{\theta}+i\, \big(\tan\theta\, F_{\phi_1}-\cot\theta\, F_{\phi_2} \big)\right]\, {\bf E}^2 +
\frac{1}{\sinh\rho}\left[ F_{\theta}-i\, \big(\tan\theta\, F_{\phi_1}-\cot\theta\, F_{\phi_2} \big)\right]\, 
\overline{\bf E^2}\cr
+\left[F_{\alpha}-i \big(\cot\alpha\, F_{\xi_1} - \tan\alpha(F_{\xi_2}+F_{\xi_3}) \big) \right]{\bf E}^5
+\left[F_{\alpha}+i\big(\cot\alpha\, F_{\xi_1} - \tan\alpha(F_{\xi_2}+F_{\xi_3}) \big) \right]\overline{\bf E^5}\cr
+\frac{1}{\cos\alpha}\left[F_{\beta}-i\big(\cot\beta\, F_{\xi_2}-\tan\beta\, F_{\xi_3})\big)\right]{\bf E}^6 
+\frac{1}{\cos\alpha}\left[F_{\beta}+i\big(\cot\beta\, F_{\xi_2}-\tan\beta\, F_{\xi_3})\big)\right]\overline{\bf E^6}\cr
 +\left[\sum_{i=0,1,2} F_{\phi_i} + \sum_{i=1,2,3} F_{\xi_i}\right]\big({\mathfrak e}^0+{\mathfrak e}^9\big)+\left[\sum_{i=0,1,2}F_{\phi_i} - \sum_{i=1,2,3} F_{\xi_i}\right]\big({\mathfrak e}^0-{\mathfrak e}^9\big) =0
\,.
\end{multline}
Here we have defined $r=l \, \sinh\rho\,$. There are two other equations for the other two constraint functions as a result of which the pullback of three of the one-forms can be eliminated in favour of the remaining ones before substituting into the BPS equations. 

For simplicity we will assume that the dual-giant under consideration is such that the following pullback conditions are trivially satisfied:
\begin{equation}\label{e5e6}
{\bf E^5} = {\bf E^6} = 0 \,.
\end{equation}
This basically reduces to the equations
\begin{align}
\dot{\alpha} = \dot{\beta} = 0,~~~  \dot{\xi_1} = \dot{\xi_2} = \dot{\xi_3} 
\end{align}
which are part of the BPS equations for the $(J_1, J_2, J_3)$ dual-giant discussed in the previous subsection. The solution to the equations \eqref{e5e6} is $\alpha = \alpha^{(0)}$\!, $\beta = \beta^{(0)}$\!, $\xi_{ij} = \xi_{ij}^{(0)}$. The constants  $\{\alpha^{(0)}, \beta^{(0)}, \xi_{12}^{(0)}, \xi_{13}^{(0)}\}$ parametrize the points on the base $\mathbb{CP}^2$ of $S^5$ which in turn parametrize the relevant maximal circles on $S^5$. We will choose our dual-giants to have only {\it one} angular momentum quantum number in the $S^5$. This amounts to fixing the parameters $\{\alpha^{(0)}, \beta^{(0)}, \xi_{12}^{(0)}, \xi_{13}^{(0)}\}$ which picks a unique maximal circle on $S^5$. Since these are two one-form conditions, we have already used two of the three constraint equations. Therefore one can substitute \eqref{e5e6} into \eqref{dFeqn}, solve for $({\mathfrak e}^0+{\mathfrak e}^9)$ and substitute into the BPS equations in \eqref{fulldggeqns}. 

The coefficients of the $4$-forms that do not vanish due to the BPS conditions lead to non-trivial constraints on the function $F$. In our case, this leads to the equations (these turn out to be simply the coefficients of the anti-holomorphic one-forms in \eqref{dFeqn}):
\begin{align}\label{dfeqns}
F_{\rho} + i\, \big(\tanh\rho\, F_{\phi_0} + \coth\rho\, (F_{\phi_1}+F_{\phi_2})\big) & = 0 \cr
F_{\theta}-i\, \big(\tan\theta\, F_{\phi_1}-\cot\theta\, F_{\phi_2}) &= 0 \cr
\sum_{i=0,1,2}F_{\phi_i} - \sum_{i=1,2,3} F_{\xi_i} &= 0 \,.
\end{align}
The solutions to the first two equations is given by
\begin{equation}\label{soln}
F = \sum_{m_0,m_1,m_2} C_{m_0,m_1,m_2}(\alpha,\beta,\xi_i)\, (\cosh\rho)^{m_0} (\sinh\rho\cos\theta)^{m_1} (\sinh\rho\sin\theta)^{m_2} e^{i\, m_0\phi_0+i\, m_1\phi_1+i\, m_2\phi_2} = 0 \,.
\end{equation}
The last equation requires that 
\begin{equation}\label{lasteqn}
\sum_{i=1,2,3} \partial_{\xi_i}C_{m_0,m_1,m_2} = i\, (m_0+m_1+m_2) C_{m_0,m_1,m_2} \,.
\end{equation}
The solution is most easily written in terms of the complex variables
\begin{equation}
\Phi_0 = l \, \cosh\rho\, e^{i\phi_0} \qquad
\Phi_1 = l\, \sinh\rho\, \cos\theta\, e^{i\phi_1} \qquad
\Phi_2 = l\, \sinh\rho\, \sin\theta\, e^{i\phi_2} \,.
\end{equation}
These are well-defined coordinates on $AdS_5$ that lead to the Bergmann form of the metric. The $\Phi_i$ satisfy the relation
\begin{equation}
-|\Phi_0|^2 + |\Phi_1|^2 + |\Phi_2|^2 = -l^2\,.
\end{equation}
In terms of these coordinates, any holomorphic function $F(\Phi_0, \Phi_1, \Phi_2)$ satisfies the first two differential equations. However, we still have to solve for the last constraint \eqref{lasteqn}. To do so, we first introduce analogous complex coordinates on the $S^5$ part of the metric as follows:
\begin{equation}
Z_1 = l \, \sin\alpha\, e^{i\xi_1} \qquad
Z_2 = l \, \cos\alpha\, \sin\beta\, e^{i\xi_2} \qquad
Z_3 = l \, \cos\alpha\, \cos\beta\, e^{i\xi_3} \,.
\end{equation}
such that
\begin{equation}
|Z_1|^2 + |Z_2|^2 + |Z_3|^2 = l^2 \,.
\end{equation}
In terms of these variables, one can check that the $\alpha=\frac{\pi}{2}$ solution to the \eqref{e5e6} is given by 
\begin{equation}
\label{halfbpsdgg1}
Z_2/Z_1 = Z_3/Z_1 = 0 \,. 
\end{equation}
Since $Z_1 \ne 0$, it follows that $Z_2$ and $Z_3$ are set to zero while $Z_1$ is just a phase ($\alpha=\frac{\pi}{2}$). The solution to \eqref{lasteqn} is therefore given by
\begin{equation}
C_{m_0,m_1,m_2} = e^{i(m_0+m_1+m_2) \xi_1} \,.
\end{equation}
Using the complex variables $\Phi_i$ and $Z_i$, one can therefore rewrite the full solution to the BPS equations in the form of the three equations
\begin{equation}
\label{halfbpsdgg2}
F(Z_1\Phi_0, Z_1\Phi_1, Z_1\Phi_2) = 0\quad\text{and}\quad  Z_2/Z_1 =  Z_3/Z_1 = 0 \,. 
\end{equation}

However this is not yet the final set of dual-giant solutions with $(S_1, S_2, J_1)$ charges we are after. We need to impose the condition that at a given fixed world-volume time the dual-giant is a point on the $S^5$. For this the spatial section of the brane at a fixed world-volume time has to be a 3-dimensional space-like surface in $AdS_5$. This is not always the case for all of the solutions given above in \eqref{halfbpsdgg2}. Even though the solutions which are not of this type might be interesting on their own, we would like to impose this condition by hand. Such solutions can be written as the following one-parameter set of 3-surfaces
\begin{equation}
\label{wobblingdgg}
F(\Phi_0 e^{-i\frac{\tau}{l}}, \Phi_1 e^{-i\frac{\tau}{l}}, \Phi_2 e^{-i\frac{\tau}{l}}) = 0, ~~~ e^{i\frac{\tau}{l}} Z_i = Z_i^{(0)} ~~ \hbox{for} ~~ i = 1, 2, 3
\end{equation}
intersected with $|\Phi_0|^2 - |\Phi_1|^2 - |\Phi_2|^2 = l^2$ and $|Z_1|^2 + |Z_2|^2 + |Z_3|^2 = l^2$ along with the condition that the intersection of $F(\Phi_0, \Phi_1, \Phi_2) = 0$  with $|\Phi_0|^2 - |\Phi_1|^2 - |\Phi_2|^2 = l^2$ is space-like. We call these solutions the ``wobbling dual-giants" \cite{ms}. Choosing $Z_2^{(0)} = Z_3^{(0)} = 0$ is the $(S_1, S_2, J_1)$ dual-giant which will be explored further later on. We note that the dual-giants in \eqref{wobblingdgg} are not the most general ones though they carry all five charges $(S_1, S_2, J_1, J_2, J_3)$. 

\vskip .5cm
\noindent{\underline{\bf{A 1/2-BPS $(S_1, S_2, J_1, J_2, J_3)$ dual-giant}}}
\vskip .5cm

 The simplest case of the single dual-giant graviton in \eqref{mssoln1} is given by the equations
\be
\label{dggofms}
\Phi_0 Z_1 = d_1, \qquad \Phi_0 Z_2 = d_2 \qquad \text{and} \qquad \Phi_0 Z_3 = d_3 \,.
\ee
The complex parameters $d_i$ can be easily written in terms of those in \eqref{mssoln1}. A simple but interesting generalization of this single dual-giant is obtained from the one in \eqref{dggofms} by rotating $(\Phi_0, \Phi_1, \Phi_2)$ with an $\frac{SU(1,2)}{U(2)}$ matrix. A general element of $\frac{SU(1,2)}{U(2)}$ is uniquely specified by three complex numbers $(c_0, c_1, c_2)$ such that 
\be
|c_0|^2 - |c_1|^2 - |c_2|^2 = 1\qquad \text{and}\qquad c_0 c_1 c_2 \in \mathbb{R}\,. 
\ee
For such a matrix $(c_0, c_1, c_2)$ make up the first row and takes $\Phi_0$ into $c_0 \Phi_0 + c_1 \Phi_1 + c_2 \Phi_2$. Such a rotation of \eqref{dggofms} gives
\begin{equation}
\label{genlindgg}
(c_0 \Phi_0 + c_1 \Phi_1 + c_2 \Phi_2) Z_1 = d_1\,, \quad (c_0 \Phi_0 + c_1 \Phi_1 + c_2 \Phi_2) Z_2 = d_2\,, \quad (c_0 \Phi_0 + c_1 \Phi_1 + c_2 \Phi_2) Z_3 = d_3 \,. 
\end{equation}
This solution has five independent complex parameters and generically has just 2 supersymmetries in  common over the five parameter space, even though at a given point on the parameter space the dual-giant is 1/2-BPS.

\subsection{Giant solutions}
We now turn to exhibiting some important known solutions to the BPS equations in \eqref{fullggeqns} for giant gravitons. Again as a check of our BPS equations \eqref{fullggeqns} describing the giant gravitons we will first show how to recover the giant graviton solutions of \cite{adkss, cs, gmnvs}. 

\subsubsection{$(S_1, S_2, J_1)$ giant}
For this we choose the static gauge $\phi_0 = \tau$, $\beta = \sigma_1$, $\xi_2 =\sigma_2$, $\xi_3 = \sigma_3$ and treat the remaining coordinates $r, \theta, \phi_1, \phi_1, \alpha, \xi_1$ as functions of $\tau$. Then the pull-backs of the space-time frame read
\begin{align}
\mathfrak{e}^0 &= [l V - \frac{r^2}{l} (\cos^2\theta  \, \dot \phi_1 + \sin^2\theta \, \dot \phi_2)] d\tau, ~~ \mathfrak{e}^1 = \frac{\dot r}{V^{1/2}} d\tau, ~~ \mathfrak{e}^2 = r \dot \theta \, d\tau, \cr
\mathfrak{e}^3 &= r V^{1/2} [ 1- \cos^2\theta \, \dot \phi_1 - \sin^2\theta \, \dot \phi_2] d\tau, ~~ \mathfrak{e}^4 = r \cos\theta \sin\theta \, (\dot \phi_1 - \dot \phi_2) d\tau, \cr
\mathfrak{e}^5 &= l \dot \alpha \, d\tau, ~~ \mathfrak{e}^6 = l \cos\alpha \, d\sigma_1, ~~ \mathfrak{e}^7 = l \cos\alpha \sin\alpha (\dot \xi_1 \, d\tau - \sin^2 \sigma_1 \, d\sigma_2 - \cos^2\sigma_1 \, d\sigma_3), \cr
\mathfrak{e}^8 &= l \cos\alpha \cos\beta \sin\beta \, (d\sigma_2 - d\sigma_3), ~~ \mathfrak{e}^9 = l[\sin^2\alpha \, \dot\xi_1 \, d\tau + \cos^2\alpha (\sin^2\beta \, d\sigma_2 + \cos^2\beta \,d\sigma_3)]. \nonumber
\end{align}
Substituting these into the last of the equations \eqref{fullggeqns} requires us to put $\dot \alpha = 0$ which means $\mathfrak{e}^5 = 0$. Using this it follows that the first of \eqref{fullggeqns} is satisfied identically. Then the equations $(\mathfrak{e}^0 + \mathfrak{e}^9) \wedge \mathfrak{\omega} \wedge \{{\bf E}^1, {\bf E}^2\} = 0$ in the third line of \eqref{fullggeqns} can be seen to be equivalent to
\begin{equation}
\dot r = \dot \theta = 0, ~~ \dot \phi_1 = \dot \phi_2 = \dot \phi_0 = 1.
\end{equation}
Using the equations $(\mathfrak{e}^0 + \mathfrak{e}^9) \wedge \mathfrak{\omega} \wedge \{{\bf E}^5, {\bf E}^6\} = 0$ we find that $\dot \xi_1 = -1$. It is simple to verify that the equations in the second line of \eqref{fullggeqns} are also satisfied. Thus we recover the equations derived in \cite{gmnvs} for these $(S_1, S_2, J_1)$ giants. The solution to these equations can be written as
\begin{equation}
\label{mssoln2}
r = r^{(0)}\,,\quad \theta = \theta^{(0)}\,, \quad \phi_1 = \phi_1^{(0)} + \tau\,,\quad \phi_2 = \phi_2^{(0)} + \tau\,, \quad  \alpha = \alpha^{(0)}\,, \quad \xi_1 = \xi_1^{(0)} - \tau
\end{equation}
with the six parameters $\{r^{(0)}, \theta^{(0)},  \phi_1^{(0)}, \phi_2^{(0)}, \alpha^{(0)}, \xi_1^{(0)} \}$ as in \cite{gmnvs}.

\subsubsection{$(J_1, J_2, J_3)$ Mikhailov giants}

The solutions analogous to the wobbling dual-giants of the BPS equations for the giant gravitons lead to the well known Mikhailov solutions \cite{mikhailov}. Let us derive this explicitly. In the differential constraint \eqref{dFeqn}, we assume now that the following pullback conditions are trivially satisfied:
\begin{equation}\label{pullbacke1e2}
{\bf E^1} = {\bf E^2} = 0 \,. 
\end{equation} 
Repeating the procedure followed for the dual-giants lead to the differential equations
\begin{align}\label{giantdf}
F_{\alpha}+i\big(\cot\alpha\, F_{\xi_1} - \tan\alpha(F_{\xi_2}+F_{\xi_3}) \big) &=0 \cr
F_{\beta}+i\big(\cot\beta\, F_{\xi_2}-\tan\beta\, F_{\xi_3}\big) &=0 \cr
\sum_{i=0,1,2} F_{\phi_i} - \sum_{i=1,2,3} F_{\xi_i}\ &= 0 \,.
\end{align}
The pullback conditions \eqref{pullbacke1e2} are solved by 
\begin{equation}
\Phi_1/\Phi_0 =  \Phi_2/\Phi_0 = 0 \,.
\end{equation}
Since $\Phi_0 \ne 0$, this fixes $\Phi_0$ to be a pure phase. Using this, the equations in \eqref{giantdf} are solved by 
\begin{equation}
F = \sum_{m_1,m_2,m_3} D_{m_1,m_2,m_3}(\phi_0)\, (\sin\alpha)^{m_1} (\cos\alpha\sin\beta)^{m_2} (\cos\alpha\cos\beta)^{m_3} e^{i\, m_1\xi_1+i\, m_2\xi_2+i\, m_3\xi_3} =0 \,,
\end{equation}
where
\begin{equation}
D_{m_1,m_2,m_3}(\phi_0) = e^{i(m_1+m_2+m_3)\phi_0} \,.
\end{equation}
Rewriting this in terms of the complex variables $Z_i$ and $\Phi_i$, we find that the solution to the differential constraint is a holomorphic function of the form
\begin{equation}
F(\Phi_0 Z_1,\Phi_0 Z_2, \Phi_0 Z_3) = 0, ~~ \Phi_1/\Phi_0 =  \Phi_2/\Phi_0 = 0 \, .
\end{equation}
This is precisely Mikhailov's solution since $\Phi_0$ is just given by the phase $e^{i\frac{t}{l}}$. A simple generalization of Mikhailov giants is obtained by letting them move along a generic time-like geodesic in $AdS_5$ used in \cite{gmnvs}. These can be written as
\begin{equation}
\label{mikhailov2}
F(e^{i\frac{\tau}{l}} Z_1, e^{i\frac{\tau}{l}} Z_2, e^{i\frac{\tau}{l}} Z_3) = 0, ~~~ e^{-i\frac{\tau}{l}} \Phi_a = \Phi_a^{(0)}
\end{equation}
intersected with $AdS_5 \times S^5$ where $\Phi_a^{(0)}$ are constants. We do not expect these to be the most general giants either even though they carry all charges $(S_1, S_2, J_1, J_2, J_3)$.

\vskip .5cm
\noindent{\underline{\bf{ A 1/2-BPS $(S_1, S_2, J_1, J_2, J_3)$ giant}}}
\vskip .5cm

The simplest example of the giants in \eqref{mikhailov2} are those in \eqref{mssoln2} which can be written as
\begin{equation}
\label{ggofms}
\Phi_0 Z_1 = c_0\,, \qquad  \Phi_1 Z_1 = c_1\,, \quad  \Phi_2 Z_1 = c_2 \,. 
\end{equation}
The parameters have to satisfy $|c_0|^2 - |c_1|^2 - |c_2|^2 \ge 0$. These carry non-zero charges $(S_1, S_2, J_1)$. They are half-BPS at any point in their parameter space but are only guaranteed to share (at least) four supersymmetries among them as we move over the parameter space.

As for the dual-giants we can generalize these further by adding four more parameters into a set of 5 (complex) dimensional space of solutions by rotating the ones in \eqref{ggofms} by a matrix in $\frac{SU(3)}{U(2)}$. Notice that one can uniquely specify a matrix in $\frac{SU(3)}{U(2)}$ by three complex numbers $(d_1, d_2, d_3)$ with the conditions 
\be
|d_1|^2 + |d_2|^2 + |d_3|^2 = 1\qquad \text{and}\qquad d_1 d_2 d_3 \in \mathbb{R}\,. 
\ee
This takes $Z_1$ into $d_1 Z_1 + d_2 Z_2 + d_3 Z_3$ so that the solution \eqref{ggofms} becomes
\begin{equation}\label{genlingg}
\Phi_0 (d_1 Z_1 + d_2 Z_2 + d_3 Z_3) = c_0, ~~ \Phi_1 (d_1 Z_1 + d_2 Z_2 + d_3 Z_3) = c_1, ~~ \Phi_2 (d_1 Z_1 + d_2 Z_2 + d_3 Z_3) = c_2. 
\end{equation}
This solution represents a single giant still that carries all charges $(S_1, S_2, J_1, J_2, J_3)$. Again at a given point in this parameter space the solution is half-BPS. However as one varies over the parameter space they share two (or more) supersymmetries among them.

\vskip .5cm
\noindent{\underline{\bf{Relation to Kim-Lee equations}}}
\vskip .5cm
 We would now like to make a general observation on the general 1/16-BPS solutions to the BPS equations. The BPS equations lead to the solution that  we restrict to holomorphic functions of the $\Phi_i$ and $Z_j$. However, there is one addition constraint, given by the last of the equations in \eqref{dfeqns}. In terms of the $\Phi_i$ and the $Z_j$, this can be rerwritten as 
\be
\sum_{i=0,1,2}\left(\Phi_i \frac{\partial F}{\partial\Phi_i} - \bar\Phi_i \frac{\partial F}{\partial\bar\Phi_i}\right) - \left(\sum_{i=1,2,3} Z_i\frac{\partial F}{\partial Z_i}- \bar Z_i \frac{\partial F}{\partial\bar Z_i}\right)=0\,.
\ee
If we restrict to holomorphic functions $F(\Phi_i, Z_j)$, as demanded by the remaining BPS equations, this equation reduces to
\be
\sum_{i=0,1,2}\Phi_i \frac{\partial F}{\partial\Phi_i} - \sum_{i=1,2,3} Z_i\frac{\partial F}{\partial Z_i} =0\,.
\ee
Thus, we recover the result derived in \cite{kimlee2} that the vector $(\Phi_0, \Phi_1, \Phi_2, -Z_1, -Z_2, -Z_3)$ be tangential to the holomorphic surface in $\mathbb{C}^{1,2}\times \mathbb{C}^3$, whose intersection with $AdS_5 \times S^5$ gives the world-volume of the giant graviton.  

\subsection{Charges}\label{charges}

In this section, we continue our analysis of the classical solutions we have described so far and obtain the momentum densities and associated charges corresponding to those solutions. We will restrict attention to dual-giants in what follows; the discussion for the Mikhailov giants proceeds along very similar lines. We work with the Lagrangian 
\be
\cal{L} = \cal{L}_{DBI} + \cal{L}_{WZ} \,,
\ee
where $\cal{L}_{DBI}$ refers to the Dirac-Born-Infeld action 
\begin{equation}
{\cal L}_{DBI} = -T_{D3} \, \sqrt{-\det_{i,j} h_{ij}} \,.
\end{equation}
Here, $T_{D3}$ is the tension of the D3-brane $T_{D3} = \frac{1}{(2\pi)^3 {\alpha'}^2 g_s}\,$, which, using the relation $4\pi g_s N = \frac{l^4}{{\alpha'}^2}$, can be written as $T_{D3} = \frac{N}{2\pi^2 l^4}$. As the relevant part of the 4-form is $C^{(4)} = - \tanh \rho \, e^{0234}$, the Wess-Zumino part of the lagrangian is 
\begin{equation}
{\cal L}_{WZ} = -T_{D_3} \, \tanh \rho \, {\mathfrak e}^{0234}.
\end{equation}
We will compute the general expression for the momentum densities in terms of $3$-forms by taking derivatives with respect to the vielbeins:
\be
p_{a} = \frac{\partial\cal{L}}{\partial \mathfrak{e}^a} \,.
\ee
Once these are computed, the momentum densities are obtained by using
\be
p_{\mu} = e^{a}_{\mu}\, p_a \,.
\ee
The $p_a$ are written as sums of $3$-forms whose coefficients are constrained by the BPS equations. We refer the reader to Appendix \ref{dualgiantcharges} for the details of the computation. A straightforward computation leads to the following expressions for the momentum densities of dual-giants. 
\begin{align}\label{momdens}
p_{1} &=\frac{N}{2\pi^2l^4}(\mathfrak{e}^{124}-\mathfrak{e}^{093})
\cr
p_2 &= -\frac{N}{2\pi^2 l^4}(\mathfrak{e}^{094}+\mathfrak{e}^{123})+ \frac{N}{2\pi^2l^4}\tanh\rho\, \mathfrak{e}^{034}\cr
p_{3}&=\frac{N}{2\pi^2l^4}(\mathfrak{e}^{091}-\mathfrak{e}^{234})-\frac{N}{2\pi^2l^4}\tanh\rho\, \mathfrak{e}^{024}
\cr
p_4&=\frac{N}{2\pi^2l^4}(\mathfrak{e}^{092}+\mathfrak{e}^{134})+\frac{N}{2\pi^2l^4}\tanh\rho\, \mathfrak{e}^{023} \cr
p_9&= \frac{N}{2\pi^2l^4}\left[\mathfrak{e}^0\wedge(\mathfrak{e}^{13}+\mathfrak{e}^{24})-\left\vert\frac{a_0}{a_1} \right\vert^2(\mathfrak{e}^0+\mathfrak{e}^9)\wedge \mathfrak{e}^{24}\right]\cr
p_0&= \frac{N}{2\pi^2l^4}\left[-\mathfrak{e}^9\wedge(\mathfrak{e}^{13}+\mathfrak{e}^{24})-\left\vert\frac{a_0}{a_1} \right\vert^2(\mathfrak{e}^0+\mathfrak{e}^9)\wedge \mathfrak{e}^{24}\right]-\frac{N}{2\pi^2l^4}\tanh\rho\, \mathfrak{e}^{234}\,.
\end{align}
Here, $a_0$, $a_1$ and $a_2$ are defined in Appendix \ref{dualgiantcharges} and arise from the differential constraint 
\be\label{defa}
a_0 {\bf E}^0 + a_1 {\bf E}^1 + a_2 {\bf E}^2 = 0  
\ee
that follows from the polynomial equation which defines the dual-giant. The integral of these momentum densities over the spatial part of the D-brane world-volume gives us the conserved charges carried by the D-brane. We will discuss several examples in the sections below. 

\vskip .5cm
\noindent{\underline{\bf{$(J_1, 0, 0)$ dual-giants}}}: Let us apply the general formulae we have obtained to the well studied case of a single 1/2-BPS dual-giant which is described by the polynomial equations
\be
f(Y_i) = \Phi_0 Z_1 - c_0 = 0 \quad \text{and}\quad  Z_2/Z_1 =  Z_3/Z_1 = 0\,.
\ee
This leads to the differential condition \eqref{defa} with 
\be
a_0 = i\qquad a_1 = \tanh\rho \qquad \text{and} ~~ a_2 = 0 \,. 
\ee
Requiring that the dual-giant is point-like in the $S^5$ direction, the differential constraints simplify to
\be
d\rho = 0 \quad \text{and}\quad d\phi_0 = -d\xi_1=d\tau \,. 
\ee
Choosing the ansatz appropriate to the spherical dual-giant, 
\be
\theta = \sigma_1 \quad \phi_1 = \sigma_2 \quad \phi_2 = \sigma_3 \,,
\ee
we find that the non-zero one-forms, when pull-backed onto the world-volume, take the form
\begin{align}
\mathfrak{e}^0 &= \cosh^2\rho\, d\sigma_0 - \sinh^2\rho(\cos^2\sigma_1\, d\sigma_2 + \sin^2\sigma_1\, d\sigma_3)\qquad \mathfrak{e}^2 = \sinh\rho\, d\sigma_1 \qquad \mathfrak{e}^9 = -d\sigma_0 \cr  
\mathfrak{e}^3 &= \sinh\rho\, \cosh\rho\, (d\sigma_0 - \cos^2\sigma_1\, d\sigma_2 - \sin^2\sigma_1\, d\sigma_3 \qquad \mathfrak{e}^4 = \sinh\rho\, \cos\sigma_1\, \sin\sigma_1(d\sigma_2-d\sigma_3)\,.
\end{align}
Using these pull-backs, we now compute the charges associated to the dual-giant; these are computed by integrating the momentum densities in \eqref{momdens} over the spatial section spanned by $
\{\sigma_{1,2,3}\}$. The relevant spatial parts of the momentum densities take the form
\begin{align}
p_1 &= p_2 = p_4 = 0 \qquad
(p_3)_{123} = -\frac{N}{2\pi^2}\frac{\sinh^3\rho}{\cosh\rho}\sin\sigma_1\,\cos\sigma_1 \cr
(p_0)_{123} &=\frac{N}{2\pi^2}\sinh^2\rho\sin\sigma_1\cos\sigma_1\qquad
(p_9)_{123} = \frac{N}{2\pi^2}\sinh^2\rho\sin\sigma_1\cos\sigma_1   \,. 
\end{align}
Now, the physically relevant momentum densities are obtained by a linear change of variables between the vielbein $e^a$ and the differentials $dX^{\mu}$. The spatial components of the momentum densities are given by
\begin{align}
p_{r}&=p_{\theta}=0 \cr
p_{\phi_0}&= (\cosh^2\rho)\, p_{0} + (\sinh\rho\cosh\rho)\, p_{3}= \frac{N}{2\pi^2}\, \sin\sigma_1\cos\sigma_1\, \sinh^2\rho\cr
p_{\phi_1}&=-\sinh^2\rho\cos^2\theta\, p_{0}-\sinh\rho\cosh\rho\, \cos^2\theta\, p_{3} +\sinh\rho\, \sin\theta\cos\theta\, p_{4}=0\cr
p_{\phi_1}&=-\sinh^2\rho\sin^2\theta\, p_{0}-\sinh\rho\cosh\rho\, \sin^2\theta\, p_{3} -\sinh\rho\, \sin\theta\cos\theta\, p_{4}=0\cr
p_{\xi_1}&=p_9 = \frac{N}{2\pi^2}\, \sin\sigma_1\cos\sigma_1\, \sinh^2\rho\,.
\end{align}
Integrating over the spatial section, we find that the only non-zero charges are given by the energy and the angular momentum along the $\alpha=\frac{\pi}{2}$ circle of the $S^5$. They satisfy the relation
\be
E = P_{\xi_1} = N\sinh^2\rho = N( |c_0|^2-1)\,.
\ee
In the last line, we have written the momenta in terms of the variables appearing in the defining equation of the dual-giant. 

\vskip .5cm 
\noindent{\underline{\bf{$(J_1, J_2, J_3)$ dual-giant}\label{j1j2j3dgg}}}: Let us generalize a little and compute the charges of a dual-giant described by the equations
\be\label{j1j2j3dggeqn}
\Phi_0 Z_1 = d_1, \qquad \Phi_0 Z_2 = d_2, \qquad \Phi_0 Z_3 = d_3 \,.
\ee
The round $S^3$ ansatz that was used in the previous example is still valid; the only difference being that $\alpha$ and $\beta$ take arbitrary values. All the momenta $p_a = \frac{\partial \cal{L}}{\partial \mathfrak{e}^a}$ computed in that section remain the same as before. However, because $\alpha \ne \frac{\pi}{2}$ anymore, the coordinate momenta change; using the appropriate vielbeins, and integrating over the spatial sections as before, we now find the following non-zero momenta
\begin{align}
P_{\xi_1} &= \sin^2\alpha\, P_9 = N\, \sinh^2\rho\, \sin^2\alpha = N\, \left(\frac{|\vec{d}|^2-1}{|\vec{d}|^2}\right) |d_1|^2\cr
P_{\xi_2} &= \cos^2\alpha\, \sin^2\beta\, P_9 =   N\, \sinh^2\rho\, \cos^2\alpha\, \sin^2\beta = N\, \left(\frac{|\vec{d}|^2-1}{|\vec{d}|^2}\right) |d_2|^2\cr
P_{\xi_3} &= \cos^2\alpha\, \cos^2\beta\, P_9 =  N\, \sinh^2\rho\, \cos^2\alpha\, \cos^2\beta = N\, \left(\frac{|\vec{d}|^2-1}{|\vec{d}|^2}\right) |d_3|^2\,,
\end{align}
where, in the last equality, we have expressed the momenta in terms of the coefficients appearing in the defining equations. The BPS equation now reads
\be
E = P_{\xi_1} + P_{\xi_2} + P_{\xi_3} = N\, \big(|\vec{d}|^2-1\big)\,.
\ee
The general expression for the momenta that have been obtained for the dual-giants can also be similarly derived for the giant gravitons and the analogous computations carried out for the Mikhailov giants as well as the $(S_1, S_2, J_1)$ giants. Our computations match the already existing results in the literature. 

 \section{Symplectic structure for wobbling dual-giants}\label{symplectic}
 
In the rest of this paper we would like to restrict ourselves to dual-giants, more specifically, to the 1/8-BPS dual-giant configurations with charges $(S_1, S_2, J_1)$. We would like to quantize the space of these solutions and see if it reproduces the answers found in \cite{gmnvs} using the language of giant gravitions with charges $(S_1, S_2, J_1)$. In this section we would like to propose that the configuration space of wobbling dual-giants with charges $(S_1, S_2, J_1)$ is a hyperbolic version of the complex projective space, with the symplectic structure given by the K\"ahler form on $\widetilde{\mathbb{CP}}^m$. However, before we proceed further we need to discuss the issue of the upper limit on the number of dual-giants. 

\subsection{Stringy exclusion principle}

An important difference between the 1/2-BPS giants and dual-giants is the way they realize the ``stringy exclusion principle". For the giant gravitons it manifests itself as the upper limit on the angular momentum $J_1$ of any given giant, and is given by $N$ \cite{mst}. For the dual-giants it appears as the upper limit on the total number of dual-giants, once again given by $N$ \cite{gmnvs} (see also \cite{bs, nvs}). It is important to understand how to impose this condition for the more general dual-giants constructed in previous section. Here we make a concrete proposal on how to implement the stringy exclusion principle for the wobbling dual-giants. For this we will start by considering the 1/2-BPS dual-giants which are given by 
\begin{equation}
F(Z_1 \Phi_0) = 0 \quad\text{and}\quad Z_2/ Z_1 = Z_3/Z_1 = 0\,.
\end{equation}
The condition that we can have at most $N$ dual-giants can be incorporated in this language by taking $f(\Phi_0 Z_1)$ to be a polynomial of order $N$, {\it i.e.}, 
\begin{align}\label{halfbpspoly}
F(\Phi_0 Z_1) &= a_0 + a_1 \, \Phi_0 Z_1 + a_2 \, (\Phi_0 Z_1)^2 + \cdots + a_N (\Phi_0 Z_1)^N = 0.
\end{align}
This simply follows from the fact that the polynomial \eqref{halfbpspoly} can be factorized uniquely into (at most) $N$ factors. Each such factor, equated to zero, is interpreted as a single dual-giant, from which it follows that an upper limit on the degree of the polynomial bounds the number of dual-giants.  

For the more general dual-giants it is not obvious how to implement this condition as they do not have simple interpretation as a configuration of non-intersecting (distinct) dual-giants. Here we propose that for the dual-giants of \eqref{halfbpsdgg2} with $(S_1, S_2, J_1)$ quantum numbers the stringy exclusion principle is implemented by restricting the degree of the variable $\Phi_0 Z_1$ in the polynomial $F(\Phi_0 Z_1, \Phi_1 Z_1, \Phi_2 Z_1)$ to $N$:
\begin{align}\label{ssjdgg}
Z_2/Z_1 &= Z_3/Z_1 = 0, \cr
F(\Phi_0 Z_1, \Phi_1 Z_1, \Phi_2 Z_1) &= \sum_{k=0}^N (\Phi_0 Z_1)^k ~ a_k(\Phi_1 Z_1, \Phi_2 Z_1) = 0 
\end{align}
where we can further write
\begin{equation}
a_k(\Phi_1 Z_1, \Phi_2 Z_1) = \sum_{i,j=0}^\infty c_{kij} ~ (\Phi_1 Z_1)^i (\Phi_2 Z_1)^j.
\end{equation}
It clearly is consistent with the 1/2-BPS ansatz \eqref{halfbpspoly} and simply amounts to generalizing it by making the constant coefficients in \eqref{halfbpspoly} functions of $\Phi_1Z_1$ and $\Phi_2 Z_1$. However, this proposal needs further justification. We will quantize the solution set \eqref{ssjdgg} later on and check that we reproduce the partition function for $(S_1, S_2, J_1)$ giants obtained in \cite{gmnvs}. 

\subsection{Symplectic structure}\label{sympintro}

We have all the tools necessary to directly compute the symplectic structure on the configuration space of wobbling dual-giants. In \cite{bglm} the authors used the covariant methods discussed in \cite{Crnkovic, Zuckerman} to find the configuration space of Mikhailov giant gravitions. We will use these methods in what follows. The basic idea is to identify the classical phase space with the space of classical solutions $\cal{M}$. The central quantity of interest is the tangent vector at a given point in $\cal{M}$. In our case, we are considering the theory on the D3 brane, so the four world-volume coordinates $\sigma_i$ play the role of the spacetime coordinates while the embedding coordinates $x^{\mu}$ and the corresponding momenta $p_{\mu}$ play the role of fields. Given a point $(x,p)$ on $\cal{M}$, which corresponds to a classical configuration that solves the equations of motion, we will denote the tangent vectors at this point by $\delta x$ and $\delta p$. These lead to infinitesimal variations of the given classical solution that do not take it away from the space of solutions. $\delta x$ (or $\delta p$) evaluated at a given $\sigma_i$ is, of course, a number. The transformation from $\delta x$ to $\delta x(\sigma)$ therefore corresponds to a one-form on the space of classical solutions, which we denote by the same symbol $\delta x(\sigma)$. One can also make higher $p$-forms by wedging together such one-forms. 

These one-forms can be used to define a symplectic current \cite{Crnkovic}, which in turn, can be used to obtain the necessary symplectic form on phase space. Once we have computed the momenta for a given classical solution, the symplectic form on phase space is simply given by
\be
\omega = \int_{\Sigma}d^3\sigma\ \delta p_{\mu}(\sigma) \wedge \delta x^{\mu}(\sigma) \,.
\ee
We would like to compute this symplectic structure on the configuration space of wobbling dual-giants. The wobbling dual-giant solution \eqref{ssjdgg} has infinitely many complex parameters $c_{n_0 n_1 n_2}$ where $0 \le n_0 \le N$ and $0 \le n_1, n_2 < \infty$. As in \cite{bglm} for Mikhailov giants we introduce a regulator $m+1$ as the number of (arbitrarily chosen) monomials $\Phi_0^{n_0} \Phi_1^{n_1} \Phi_2^{n_2}$ that appears in the power series of $F(\Phi_0, \Phi_1, \Phi_2)$. The corresponding polynomial will have $m+1$ coefficients and multiplying them by a non-zero complex number does not change the solution. Therefore as in \cite{beasley, bglm} one expects the parameter space of the wobbling dual-giants also to be a complex projective space.  We propose that it is actually given by a hyperbolic projective space $\widetilde{\mathbb{CP}}^m$. As evidence towards this conjecture we will now compute the moduli space for two special cases of 1/8-BPS wobbling dual-giants: 
\begin{itemize}
\item{The 1/2-BPS dual-giants in \eqref{halfbpspoly}. We argue that moduli space in this case is $\widetilde{\mathbb{CP}}^N$ with each of its inhomogeneous coordinates being charged under the generator corresponding to the charge $J_1$,}
\item{The 1/8-BPS linear polynomial in \eqref{genlindgg} with $d_1 = d, d_2 = d_3 = 0$. We show that it has the moduli space $\widetilde{\mathbb{CP}}^3$ with the three inhomogeneous coordinates carrying a unit of charges $S_1$, $S_2$ and $J_1$ each.}
\end{itemize}

\subsubsection{1/2-BPS dual-giants}

We begin with the 1/2-BPS dual-giants with the non-zero $J_1$ charge. These are described by the defining equaitons
\be
f(\Phi_0 Z_1)= \sum_{k=0}^N a_k \, (\Phi_0 Z_1)^k = 0\quad \text{and}\quad  Z_2/Z_1 = Z_3/Z_1 = 0 \,.
\ee
For the linear polynomial of case, $f(\Phi_0 Z_1)= \Phi_0 Z_1 -c_0$, the phase space was computed in \cite{gmnvs}. We will rederive this result using a different method. The momentum densities for this configuration have already been computed in the previous section, the only non-zero charge comes from the momentum density along the $\xi_1$ direction
\be
p_{\xi_1} = \frac{N}{2\pi^2}\, \sinh^2\rho \sin\sigma_1 \cos\sigma_1 = \frac{N}{2\pi^2}\, (|c_0|^2-1) \sin\sigma_1 \cos\sigma_1\,,
\ee
where we have expressed $\rho$ in terms of the parameters appearing in the defining equation. Prior to computing the symplectic form $\omega$, let us compute the one-form $\theta$, whose derivative is $\omega=d\theta$. Since $p_{\rho}=0$, we get the simple expression
\be
\theta = \int_{\Sigma} p_{\xi_1} \delta\xi_1 \,.
\ee
So it remains to compute the variation $\delta\xi_1$. From the defining equation, it is not difficult to see that
\be
\delta\xi_1 = \frac{1}{2i}\left(\frac{\delta c_0}{c_0} - \frac{\bar\delta c_0}{\bar c_0}\right)\,.
\ee
After performing the integral over the spatial section of the dual-giant, $\theta$ is then given by 
\be
\theta = \frac{N}{2i} (|c_0|^2-1)\left(\frac{\delta c_0}{c_0} - \frac{\bar\delta c_0}{\bar c_0}\right)\,. 
\ee
Differentiating, we obtain the symplectic form on the configuration space of the single 1/2-BPS dual-giant:
\be\label{omegac}
\omega = -iN\, \delta\bar c_0 \wedge \delta c_0 \,.
\ee
Since $|c_0| > 1$, we see that the symplectic form coincides with that on the outside of a disk of unit radius in the complex plane. 

The configuration space of 1/2- BPS dual-giants was also computed in \cite{gmnvs}, so let us try to compare the two results. It was shown in that reference that the supersymmetry constraints made the configuration space into a reduced phase space. Using the Dirac brackets the symplectic structure was found to be 
\be
\omega = -i \, N\, \delta\bar\zeta \wedge \delta \zeta \qquad \text{where}\qquad \zeta = r_0 \, e^{i \, {\xi_1^{(0)}}}\,. 
\ee
In terms of the coordinates of $AdS_5 \times S^5$, the parameter $c_0$ is given by (we set $l=1$ in all computations from here) 
\be\label{candr}
c_0= \sqrt{r_0^2 + 1}\,  e^{i\,\xi_1^{(0)}}
\ee
where $r_0$ is the position of the dual-giant in the radial position in $AdS_5$ and $\xi_1^{(0)}$ is the position of the dual-giant in the $\xi_1$ direction at $\tau = 0$. Given this, it is easy to see that this symplectic structure, rewritten in terms of the variables $c_0$ using 
\be
\zeta = \frac{c_0}{|c_0|} \sqrt{|c_0|^2 - 1} 
\ee
remains form invariant and is given by \eqref{omegac}, with the restriction that $|c_0| > 1$. Thus we conclude that the configuration space of a single 1/2-BPS dual-giant is a copy of $\mathbb{C}^1$.

Now, let us turn to the multiple dual-giant case with 
\be
f(\Phi_0Z_1) = \prod_{i=1}^{N}(\Phi_0 Z_1 - c_0^{(i)}) = 0 \,.
\ee
The key point to note is that, in this case, one has to sum over the $N$ zeroes of $f(\Phi_0 Z_1)$. Following this prescription one finds that the full symplectic form is given by
\be
\omega=-i\, N\sum_{i=1}^{N}\delta \bar{c}_0^{(i)} \wedge \delta c_0^{(i)} \,.
\ee
Therefore the configuration space of the 1/2-BPS polynomial is the symmetrized product of the $N$ copies of the configuration space of a single 1/2-BPS dual-giant. However, in order to generalize this discussion to the 1/8-BPS dual-giants, it would be useful to give a slightly different description of the configuration space. We will now argue that this configuration space $(\mathbb{C}^1)^N/S_N$ of 1/2-BPS dual-giants can be mapped onto the hyperbolic space $\widetilde{\mathbb{CP}}^N$. 

Before we turn to this we will present a useful coordinate transformation. We will show that (i) $\mathbb{C}^m$ can be mapped onto $\widetilde{\mathbb{CP}}^m$ and (ii) the interior of the unit disc in $\mathbb{C}^m$ can be mapped onto a $\mathbb{CP}^m$ such that the standard K\"ahler form on $\mathbb{C}^m$ gets mapped onto the Fubini-Study $2$-form on the respective K\"ahler manifolds. Let us begin with the standard K\"ahler form on $\mathbb{C}^m$:
\begin{equation}\label{flatomega}
\omega = -i \sum_{i=1}^{m} \delta\bar \zeta_i \wedge \delta\zeta_i.
\end{equation}
Consider the change of variables 
\be
\label{cntocpn}
\zeta_ i = b_i\, \sqrt{f(\vert b\vert^2)} \,,
\ee
where $|b|^2 = |b_1|^2 + \cdots +|b_N|^2$. Then, we get the differential conditions
\begin{align}
\delta\zeta_i &= \delta b_i\, \sqrt{f(\vert b\vert^2)} + \frac{f'(\vert b\vert^2)}{2\sqrt{f(\vert b\vert^2)}} b_i\, \delta(\vert b\vert^2) \cr
\delta\bar \zeta_i &= \delta\bar b_i\, \sqrt{f(\vert b\vert^2)} + \frac{f'(\vert b\vert^2)}{2\sqrt{f(\vert b\vert^2)}}\bar b_i\,  \delta(\vert b\vert^2) \,. 
\end{align}
Substituting these into the symplectic form \eqref{flatomega} and using
\be
\delta(\vert b\vert^2) = \sum_{i} (b_i \delta\bar b_i + \bar b_i \delta b_i) \,, 
\ee
we get
\be\label{sympusingf}
\omega = -iN \left[f(\vert b\vert^2)\, \sum_{i} \delta b_i \wedge \delta\bar b_i + f'(\vert b\vert^2)\sum_{i,j} \bar b_i b_j \delta b_i\wedge \delta\bar b_j\right] \,. 
\ee
Consider now two choices for $f(b)$, given by 
\be\label{fmap}
f_{\pm}(\vert b\vert^2) = \frac{1}{1\pm \vert b\vert^2} 
\ee
When $f=f_+$, we see that we have obtained the K\"ahler form on the complex projective space $\mathbb{CP}^m$ while for $f=f_-$, we get the K\"ahler form on the negatively curved hyperbolic space $\widetilde{\mathbb{CP}}^m$. Moreover, from \eqref{cntocpn}, we see that it is only the unit disc in $\mathbb{C}^m$ that gets mapped onto the positively curved $\mathbb{CP}^m$ while the entire $\mathbb{C}^m$ is mapped onto $\widetilde{\mathbb{CP}}^m$ using this change of variables. 

Now we return to the 1/2-BPS dual-giant configuration space. Consider each factor $\Phi_0 = c_0^{(i)}$ in the half-BPS polynomial at a time. The coordinates $c_0^{(i)}$ are such that $|c_0^{(i)}|\ge 1$. We make the change of coordinates 
\begin{equation}
c_0^{(i)} \rightarrow \zeta_i = \frac{c_0^{(i)}}{\sqrt{1-|c_0^{(i)}|^2}}
\end{equation}
so that $\zeta_i$ is a coordinate on $\mathbb{C}^1$.  Now we introduce a regulator $|\zeta_i| \le \mathfrak{r}$ for each $\zeta_i$. We can map the remaining disc into a copy of $\mathbb{CP}^1$ using the map above with negative sign for $N=1$. Then we have the configuration space of the 1/2-BPS dual-giants, with the regulator in place, to be $(\mathbb{CP}^1)^N/S_N$ which, in turn, is equivalent to $\mathbb{CP}^N$ (see \cite{bglm} for instance). We can now use the inverse map in \eqref{cntocpn} to map the configuration space to a disc in $\mathbb{C}^N$ with the standard K\"ahler form with the size of the disc set by the regulator $\mathfrak{r}$. When we remove the regulator we end up with $\mathbb{C}^N$. We can further map this $\mathbb{C}^N$ onto $\widetilde{\mathbb{CP}}^N$ using the mapping \eqref{cntocpn}. 

\subsubsection{A single $(S_1, S_2, J_1)$ dual-giant}

We will now start from the result we have just obtained for the simple 1/2-BPS spherical dual-giant and compute the symplectic structure on the configuration space of the 6-parameter linear polynomial solution 
\begin{equation}
\label{genlindgg2}
(c_0 \Phi_0 + c_1 \Phi_1 + c_2 \Phi_2) Z_1 = d\,, ~~ Z_2/Z_1 = Z_3/Z_1 = 0 
\end{equation}
with
\be
|c_0|^2 - |c_1|^2 - |c_2|^2 = 1 \quad \text{and}\quad c_0 c_1 c_2 \in \mathbb{R} \,.
\ee
The 1-form $\theta$ for these solutions is given by
\begin{equation}\label{deftheta}
\theta = \sum_{i=0}^{2} P_{\phi_i} \, \delta \phi_i \,. 
\end{equation}
We first compute the momenta and variations in \eqref{deftheta}. The key point to note is that the general linear polynomial in \eqref{genlindgg2} can be obtained from the simple round $S^3$ dual-giant by an $SU(1,2)/U(2)$ rotation matrix. For the 1/2-BPS case defined by the equation $\Phi_0 Z_1 = d$, recall that $\theta$ is given by
\be
\theta = \frac{N}{2i} \big(|d|^2-1\big) \, \left(\frac{\delta d}{d}-\frac{\delta\bar{d}}{\bar{d}}\right) \,. 
\ee
Here, we have already integrated over the volume of the round $3$-sphere. Defining  
\be
\vec{\mathfrak{c}} = \left(\frac{c_0}{d}, \frac{c_1}{d}, \frac{c_2}{d}\right)
\ee
we find that
\be 
|\mathfrak{c}|^2 = \frac{1}{|d|^2}(|c_0|^2 - |c_1|^2 - |c_2|^2) = \frac{1}{|d|^2} \,. 
\ee
Furthermore, defining the $SU(1,2)$-invariant form $\eta^{ij} = \hbox{diag} (+1, -1, -1)$, one can easily generalize the one-form $\theta$ of the round $S^3$ dual-giant to the corresponding one-form $\theta$ in the configuration space of the linear polynomial:
\begin{equation}
\theta = -\frac{N}{2i} \left( \frac{1}{|\mathfrak{c}|^4}-\frac{1}{|\mathfrak{c}|^2} \right) \eta^{ij} \big(\bar{\mathfrak{c}_i} \, \delta\mathfrak{c}_j - \mathfrak{c}_j\, \delta\bar{\mathfrak{c}_i}\big) 
\end{equation}
Defining the new variables
\be
\lambda_i = \sqrt{\frac{1-|\vec{\mathfrak c}|^2}{|\vec{\mathfrak c}|^4}}\, \mathfrak{c}_i \quad\text{for}\quad i = 0, 1, 2\,,
\ee
and recasting the one-form $\theta$ in these variables gives
\be
\theta = \frac{iN}{2}\eta^{ij} \big(\bar{\lambda}_{i}\, \delta\lambda_{j} - \lambda_{j}\, \delta\bar{\lambda}_{i}\big) \,.
\ee 
Observe that since $|d| > 1$, the vector $\vec{\mathfrak{c}}$ always has its norm to be less than unity. This implies that 
\be
|\lambda|^2  = |\lambda_0|^2 - |\lambda_1|^2-|\lambda_2|^2 > 0 \,.
\ee
This implies that the phase space for the general linear polynomial is very simply described in the $
\lambda$-variables: it is the region inside the light-cone of $\mathbb{C}^{1,2}$ with the symplectic form 
\be
\omega = iN\big(\delta\bar{\lambda}_0 \wedge\delta\lambda_0-\delta\bar{\lambda}_1 \wedge\delta\lambda_1 -\delta\bar{\lambda}_2 \wedge\delta\lambda_2\big) \,.
\ee
The conserved charges, in these variables, are given by 
\be
\tilde Q_i = N\, |\lambda_i|^2 \quad \text{for}\quad i=0,1,2\,,
\ee
which map to $(E, S_1, S_2)$ respectively. The domain in $\mathbb{C}^{1,2}$ of interest can be mapped onto a Euclidean signature space by the following change of variables:
\begin{align}
\eta_0 = \frac{|\lambda_0|}{\lambda_0}\, |\lambda| \qquad
\eta_1 = \frac{|\lambda_0|}{\lambda_0}\, \lambda_1 \qquad
\eta_2 = \frac{|\lambda_0|}{\lambda_0}\, \lambda_2 \,.
\end{align}
In terms of the $\eta_i$, one can check that the configuration space for the linear $(S_1, S_2, J_1)$ dual-giant is simply $\mathbb{C}^3$, with the symplectic form 
\be\label{sympeta}
\omega = -iN\, \sum_{i=0}^{2}\delta\bar{\eta}_i \wedge \delta\eta_i \,.
\ee
The conserved charges, in these variables are given by
\be
Q_i =N  |\eta_i|^2 \quad\text{for} \quad i=0,1,2 \,.
\ee
and correspond to $(J_1, S_1, S_2)$ respectively. From this, one can easily read off the relevant momenta of the dual-giant corresponding to the linear polynomial and rewrite them in the original variables. We find 
\begin{align}
P_{\phi_i} &= N\, \left(\frac{1-|\vec{\mathfrak c}|^2}{|\vec{\mathfrak c}|^4}\right) |c_i|^2 \quad \text{for}\quad i=0,1,2 \cr
\sum_{i=0}^{3} \eta^{ii} P_{\phi_i} &= N \, \left(\frac{1-|\vec{\mathfrak c}|^2}{|\vec{\mathfrak c}|^2}\right) \equiv J \,.
\end{align}
Finally we can now map the configuration space of the linear polynomial into a $\widetilde{\mathbb{CP}}^3$ using the map \eqref{cntocpn} as promised. Defining
\be
\eta_i = \frac{1}{\sqrt{1-|b|^2}}\  b_i \,, 
\ee
the $b_i$ denote coordinates on $\widetilde{\mathbb{CP}}^3$. In these variables, the symplectic form in \eqref{sympeta} maps to the K\"ahler form on $\widetilde{\mathbb{CP}}^3$, as shown in equation \eqref{sympusingf}. The charges, in the $b$-variables, take the form 
\be\label{chargesb}
\frac{Q_i}{N} = \frac{|b_i|^2}{1-|b|^2} \,.  
\ee

\section{Counting wobbling dual-giants}\label{quantum}

We have derived in the previous section that the configuration space of a single dual-giant with spins $(S_1, S_2, J_1)$ is the complex projective space $\widetilde{\mathbb{CP}}^3$ with the symplectic structure given by the K\"ahler form. Similarly, for the 1/2-BPS dual-giants with $(0, 0, J_1)$, we have shown that the configuration space is given by $\widetilde{\mathbb{CP}}^N$. 

For the general case, as mentioned at the beginning of Section \ref{sympintro}, we conjecture that the configuration space is given by $\widetilde{\mathbb{CP}}^m$ with the regulator in place, in the defining polynomial: 
\be
F(\Phi_0 Z_1, \Phi_1 Z_1, \Phi_2 Z_1) = \sum_{k=0}^N \sum_{i,j=0}^{\infty} c_{ijk}\, (\Phi_0 Z_1)^k (\Phi_1 Z_1)^i\, (\Phi_2 Z_1)^j = 0 \, .
\ee
Analogous to the 1/2-BPS case discussed in Appendix \ref{quantumcptilde}, one can define new coordinates $b_{n_0 n_2 n_2}$ with $0 < n_0 \le N, n_1, n_2 \ge 0$, such that the charges of the dual-giants are given by
\begin{align}
\frac{J_1}{N} &= f (b) \sum_{n_0, n_1 n_2}  n_0 |b_{n_0 n_2 n_2}|^2 \cr 
\frac{S_i}{N} &= f (b) \sum_{n_0, n_1, n_2} n_i |b_{n_0 n_2 n_2}|^2\quad \text{for}\quad i=1, 2\,, 
\end{align}
where 
\be
f(b) = \frac{1}{(1-\sum_{n_0, n_1 n_2}  |b_{n_0 n_2 n_2}|^2)}\,. 
\ee
Note that these expressions agree with the expressions for the charges derived in \eqref{Jdefb} for the 1/2-BPS case as well as equation \eqref{chargesb} for the single dual giant described by a linear polynomial.  We also observe that the expressions for the charges we have written down, apart from the restriction on $n_0$, is similar to the expression for the $(J_1, J_2, J_3)$ charges of the Mikhailov giants discussed in \cite{bglm}.\footnote{In fact, one can map our wobbling dual-giants into Mikhailov giants by a double Wick-rotation that interchanges $AdS_5$ with $S^5$, at least over some subspaces of their respective parameter spaces.} The appearance of the weighted sum can be understood by focusing on the 1/2-BPS case. The $b_k$'s are then symmetric polynomials of the roots of the defining polynomial. Since the individual roots have a unit charge, the $k$th symmetric combination $b_k$ has charge $k$. 

It will be desirable to have a direct way to verify whether or not the moduli space of $(S_1, S_2, J_1)$ dual-giants, with the regulator $m$ in place, is given by $\widetilde{\mathbb{CP}}^m$. We will assume that it is given by a $\widetilde{\mathbb{CP}}^m$ with the standard K\"ahler form on it and proceed with its quantization. This can be done in two different ways. One is to use the inverse map in \eqref{cntocpn} to map the problem on to $\mathbb{C}^k$. Then the holomorphic quantization is immediate with the result that the Hilbert space is given by holomorphic functions of arbitrary degree. Taking the monomials 
\begin{equation}
\psi_{p_{n_0 n_1 n_2}} (\tilde b_{n_0 n_1 n_2}) := \prod_{n_0 n_1 n_2} (\tilde b_{n_0 n_1 n_2} )^{p_{n_0 n_1 n_2}}
\end{equation}
to be a basis for the Hilbert space were we used $\tilde b_{n_0n_1n_2}$ to denote the tranformed coordinates on $\mathbb{C}^k$. Then the charges of these basis states are 
\begin{equation}
J_1 = \sum_{n_0 n_1 n_2} n_0 \, p_{n_0 n_1 n_2}\,, \quad S_i = \sum_{n_0 n_1 n_2} n_i \, p_{n_0 n_1 n_2}\quad \text{for}\quad i =1,2 \,.
\end{equation}
where the sums are over $\{0 < n_0 \le N, 0 \le n_1, n_2 < \infty \}$ with the regulator $m$ in place.  When we take the regulator away the Hilbert space can be identified with that of an arbitrary number of bosons in a 3-dimensional harmonic oscillator with the level numbers of the single-particle Hilbert space labeled by $\{n_0, n_1, n_2 \}$ with the restriction $0 < n_0 \le N$. The partition function of this system can be written as
\begin{equation}
Z(q_0, q_1, q_2) = \prod_{n_0=1}^N \prod_{n_1, n_2 = 0}^\infty \frac{1}{1- q_0^{n_0} q_1^{n_2} q_2^{n_2}}\,.
\end{equation}
This matches precisely with the one obtained by counting the giants as in \cite{gmnvs} as well as the gauge theory answer of \cite{ggkm}. The second method of quantization of our problem is to directly use the quantization of the kahler manifold $\widetilde{\mathbb{CP}}^k$. This gives a Hilbert space isomorphic to the one obtained above. See Appendix  \ref{quantumcptilde} for some details of this method. 

It is amusing to note that the quantization of the wobbling dual-giants naturally gives rise to a description of the Hilbert space as that of an arbitrary number of bosons in a three dimensional harmonic oscillator, with one of the level numbers restricted to be less than $N$. This is precisely the dual description used in \cite{gmnvs} to quantize the $(S_1, S_2, J_1)$ giants. This is similar to what was observed for the Mikhailov giants, whose quantization gave the dual description in terms of $N$ bosons in a three dimensional harmonic oscillator. 

\section{Discussion}\label{discuss}

In the first part of this article, we found the BPS equations for D3-branes embedded in $AdS_5\times S^5$ that preserve two out of the full thirty-two supercharges and recovered large classes of solutions that were studied in the literature. We then focused predominantly on the dual-giants with charges $(S_1, S_2, J_1)$ and argued that their configuration space can be mapped to the hyperbolic version of the complex projective space $\widetilde{\mathbb{CP}}^m$ where $m$ is a regulator. The description of our dual-giants is different from those considered in \cite{kimlee2, ggkm} as deformations of the spherical dual-giants. We made a specific conjecture on how to implement the stringy exclusion principle. We then argued that one can recover the partition function of $(S_1, S_2, J_1)$ 1/8-BPS states computed earlier using giants \cite{gmnvs} and gauge theory \cite{ggkm}. It will be interesting to verify our conjecture on the implementation of the stringy exclusion principle from other sources. 

\vskip .5cm
In what follows, we mention some connections of our work to existing literature and point out possible avenues for future work. 

\vskip .5cm
\noindent{\underline{\bf Generalizations}}: The techniques introduced in this paper to analyze the $\kappa$-symmetry conditions and subsequent solutions are general and can prove useful in understanding the embedding of other extended objects in $AdS_5 \times S^5$. Our calculations in obtaining the BPS equations can be generalized to the case of giant gravitons in the maximally supersymmetric $AdS_{4} \times S^{7}$ or $AdS_7\times S^4$ backgrounds of M-theory. 

One can also generalize to less supersymmetric cases such as the $AdS_5 \times Y^{p,q}$ backgrounds \cite{gmsw}. One expects that the generalization of the BPS equations for giants and dual-giants in these backgrounds should be given by ones similar to (\ref{fullggeqns}, \ref{fulldggeqns}) with the K\"ahler form $\omega$ on the base manifold $\mathbb{CP}^2$ of $S^5$ replaced by the K\"ahler form on the appropriate base manifold of $Y^{p,q}$ and $e^9$ replaced by the 1-form dual to corresponding the Reeb vector, which generates translations along the fibre coordinate. It will be interesting to recover the objects in \cite{ms, bm, gn} and classify and count the D3-branes with two supercharges in these backgrounds. These objects should be relevant for the microstate counting of some of the black holes \cite{gr1} lifted to asymptotically $AdS_5 \times Y^{p,q}$ solutions using  \cite{bl}. Another case of interest with less supersymmetry is to further study probe branes in the near-horizon geometries of black holes in $AdS_5 \times S^5$ \cite{sss, ss}. Some of these topics will be discussed elsewhere.

\vskip .5cm
\noindent{\underline{\bf{Calibrations}}}: The equations that followed from imposing kappa-symmetry for D3 branes led to very simple constraints on the pull-back of $4$-forms in space-time. It is worthwhile to point out that our general analysis of the kappa-symmetry equations agrees with results obtained using rather different techniques in \cite{h-js}. Irrespective of whether we are considering giants or dual-giants, the space-time volume-form on the world-volume of the D3 brane we have derived, after imposing the supersymmetry conditions is given by 
\be
\text{dvol}_{4} = \big\vert e^{09} \wedge (\tilde{\bf \omega}-{\bf \omega})\big\vert \,.
\ee
The embedding of the D3 brane is such that the spatial part of the world-volume turns out to be
\begin{align}
\text{dvol}_{3}&= \left\vert e^9 \wedge {\bf \omega}\right\vert \qquad \text{for giants, and}\cr
\text{dvol}_{3}&= \left\vert e^0 \wedge \tilde{\bf \omega}\right\vert \qquad \text{for dual-giants}\,,
\end{align}
where $\tilde{\bf \omega}$ and ${\bf \omega}$ are the respective K\"ahler forms on $\widetilde{\mathbb{CP}}^2$ and $\mathbb{CP}^2$ respectively. This precisely coincides with the calibration forms obtained for the giants and dual-giants in \cite{h-js} and is a useful check of our supersymmetry analysis.

\vskip .5cm
\noindent{\underline{\bf Counting and plethystics}}: As was mentioned in the introduction, there are dual descriptions for the 1/8-BPS giants/dual-giants with a given set of quantum numbers. It turns out that one of the two descriptions is a ``free" system while the other describes an interacting one. For instance, for the 1/8-BPS states with quantum numbers $(J_1, J_2, J_3)$, the description in terms of the Mikhailov giants is the interacting one and the subsequent quantization of the multi-giant configuration space required fairly sophisticated techniques \cite{bglm}. This is in contrast with the elementary description of the $(J_1, J_2, J_3)$ dual-giants in \cite{gmnvs}. Similarly, the wobbling giants, also introduced in \cite{gmnvs} describe a free system while the wobbling dual-giants, studied in the present work, describes an interacting system. 

We would like to make the observation that in the case when the dual free description is available, the plethystic  techniques of \cite{ami1, ami2} can be used to obtain the multi-particle partition function, given the single particle partition function. For instance, the single particle configuration space of the $(S_1, S_2, J_1)$ giants is given by $\mathbb{C}^2 \times \mathbb{D}$, where $\mathbb{D}$ is the unit disc in $\mathbb{C}$. The single giant partition function is given by
\be
Z_{1} = \frac{1+q_0+\ldots + q_0^{N}}{(1-q_1)(1-q_2)} \,.
\ee
This is what is denoted $f_{\infty} = g_1$ in \cite{ami1}. Using the plethystic exponential, one can now easily obtain the multi-giant partition function 
\be
Z_{\infty} = {\rm PE}\big[Z_1\big] =  \prod_{n_0=1}^N \prod_{n_1, n_2 = 0}^\infty \frac{1}{1- q_0^{n_0} q_1^{n_2} q_2^{n_2}} \,.
\ee
Similarly, the generating function $g(\nu, t)$ introduced in \cite{ami2} can be used to obtain the grand canonical partition function for the $(J_1, J_2, J_3)$ dual-giants derived in \cite{gmnvs} using the fact that the partition function of a single dual-giant is the same as that of a single particle in a three dimensional harmonic oscillator. However, such techniques prove inadequate when applied to the interacting description of the BPS states. 

\vskip .5cm
There remain some important questions that one has to answer in the program of counting giants and dual-giants, we point out a few of these in what follows.

\vskip .5cm
\noindent{\underline{\bf{1/16-BPS states}}}: Recently \cite{ggkm} has made some progress towards counting 1/16-BPS giants building upon the work of \cite{kimlee2}. In particular, states involving scalar fields and covariant derivatives have been enumerated in \cite{ggkm}. Ideally, one would like to compute the 1/16-BPS states in $AdS_5 \times S^5$ by counting giants and/or dual-giants and recover the results of \cite{ggkm}. Although this is beyond the scope of this paper, we can, using the techniques in our paper, obtain the configuration space and symplectic form for a class of dual-giants and giants, described respectively by the equations \eqref{genlindgg} and \eqref{genlingg}. Although these are 1/2-BPS states at a given point in parameter space, they preserve only 2 supersymmetries over the full parameter space.

The methods are very similar to the ones we employed for the $(S_1, S_2, J_1)$ dual-giants, so we will only quote the answers. For the dual-giants described by \eqref{genlindgg}, we find that the 10-dimensional configuration space of a single dual-giant is given by a warped product of $\mathbb{C}^3 \times \widetilde{\mathbb{CP}}^2$. Similarly, for the single giant of \eqref{genlingg}, we find that the configuration space is a warped product of $\mathbb{CP}^3 \times \widetilde{\mathbb{CP}}^2$. It remains to be seen whether one can try and generalize this and recover the results of \cite{ggkm}. 

\vskip .5cm
\noindent{\underline{\bf EM waves}}: It has been shown in \cite{kimlee1} that there are giant gravitons with world-volume electro-magnetic fields turned on. Similarly there are supersymmetric dual-giant gravitons with electro-magnetic fields \cite{ss}. Their dual descriptions are as yet unknown. The dual description of giants with EM fields are not given by the dual-giants with EM fields. The giants with EM waves preserve the $SO(4)$ symmetry coming from the $AdS_5$ part of the geometry, just as those giants without the EM fields, whereas the dual-giants with EM fields break this symmetry. This suggests that one should consider higher dimensional branes such as D5-branes which preserve the same isometry as the configuration of giants or dual-giants with EM fields. Our techniques should be useful in analyzing, generalizing and classifying the solutions of \cite{kimlee1, ss} as well as their duals \cite{asnext}. 

\section*{Acknowledgements}

We would like to thank Ghanashyam Date, Eleonora Dell'Aquila, Jerome Gauntlett and Jaume Gomis for helpful discussions. Research at the Perimeter Institute is supported by the Government of Canada through Industry Canada and by the Province of Ontario through the Ministry of Research and Innovation. 

\begin{appendix}

\section{Computation of charges}\label{dualgiantcharges} 

In order to get compact expressions for the momentum densities, it will be easiest to work with the complex and real forms introduced earlier in \eqref{cplxforms} and \eqref{realforms}. For dual-giants, recall that the pullback onto the world-volume of the following forms are zero:
\be
\bf{E}^5={\bf E}^6 = 0
\ee
In terms of the remaining combinations of $1$-forms, the determinant of the induced metric on the world-volume is given by 
\begin{multline}
-\det h=\frac{1}{4}|{\bf E}^{0\bar012}|^2 + \frac{1}{16}|{\bf E}^{0\bar0}({\bf E^{1\bar1}+E^{2\bar2}})|^2 \cr
+\frac{1}{4}\left[|\mathfrak{e}^0\wedge {\bf E}^{12\bar2}|^2+|e^0\wedge {\bf E}^{21\bar1}|^2-|\mathfrak{e}^9\wedge {\bf E}^{12\bar2}|^2-|\mathfrak{e}^9\wedge \bf{E}^{21\bar1}|^2\right]-|{\bf E}^{1\bar12\bar2}|^2 \,.
\end{multline}
Classically, the D3-brane with spins $(S_1, S_2, J_1)$ is described by the equation
\be
f(Y_0, Y_1, Y_2) = 0 \quad\text{where}\quad Y_k = \Phi_k Z_1\quad k\in \{1,2,3\} \,.
\ee
This leads to conditions on the pull-backs of the bulk 1-forms. We have the relation
\begin{multline}
[\tanh \rho \, f_0 Y_0 + \coth \rho \, (f_1 Y_1 + f_2 Y_2) ] (\mathfrak{e}^1 - i\mathfrak{e}^3) \\
 - \frac{1}{\sinh\rho} [\tan\theta \, Y_1 f_1 - \cot\theta \, Y_2 f_2] (\mathfrak{e}^2-i\mathfrak{e}^4) \\
+ i [ Y_0 f_0 + Y_1 f_1 + Y_2 f_2] (\mathfrak{e}^0 + \mathfrak{e}^9)  = 0.
\end{multline}
This is just a rewriting of \eqref{dFeqn} in terms of $Y_k$-derivatives. Let us write this equation as 
\be
a_0 {\bf E}^0  +a_1 {\bf E}^1 + a_2 {\bf E}^2 = 0\,. 
\ee
Its conjugate then reads 
\be
\bar a_0 {\bf E}^0 + \bar a_1 {\bf \bar E}^1 + \bar a_2 \bf{\bar E}^2 = 0\,.
\ee 
since ${\bf E}^0$ is real (note that ${\bf \bar E}^0 = \mathfrak{e}^0 - \mathfrak{e}^9$ is not the complex conjugate of $\bf E^0$). Then it is easy to see that the pull-backs of the following 4-forms constructed out of $\{{\bf E}^0, {\bf \bar E}^0, {\bf E}^1, {\bf \bar{E}}^1, \bf{E}^2, {\bf \bar{E}}^2 \}$ vanish identically: 
\be
{\bf E}^{0\bar012} = {\bf E}^{0\bar0 \bar 1 \bar 2}={\bf E}^{01\bar 1 2}={\bf E}^{01\bar 1\bar 2}={\bf E}^{012\bar2}={\bf E}^{0\bar 12\bar 2}={\bf E}^{1\bar 1 2\bar 2}=0\,. 
\ee
All the remaining eight 4-forms can be written in terms of just one of them as follows: 
\begin{multline}
{\bf E}^{0\bar01\bar2} = -\frac{\bar{a_{1}}}{\bar{a_2}}\, {\bf E}^{0\bar01\bar1}\,, \quad {\bf E}^{0\bar0\bar12}=\frac{a_1}{a_2}\, {\bf E}^{0\bar01\bar1}\,, \quad {\bf E}^{0\bar02\bar2}=\left\vert\frac{a_1}{a_2}\right\vert^2\, {\bf E}^{0\bar01\bar1} \,,\quad {\bf E}^{\bar01\bar1\bar2} = \frac{\bar{a_0}}{\bar{a_2}}\, {\bf E}^{0\bar01\bar1}\cr
{\bf E}^{\bar01\bar12}=\frac{a_0}{a_2}\, {\bf E}^{0\bar01\bar1}\,,\quad {\bf E}^{\bar0\bar12\bar2}=\frac{a_1\bar{a_0}}{|a_2|^2}\, {\bf E}^{0\bar01\bar1}\,, \quad {\bf E}^{\bar012\bar2}= \frac{a_0\bar{a_1}}{|a_2|^2}\, {\bf E}^{0\bar01\bar1}\,. 
\end{multline}
We have similar relations among the 3-forms. Out of all the possible twenty 3-forms that can be written out of the available 1-forms only ${\bf E}^{012}$ and ${\bf E}^{0\bar 1 \bar 2}$ vanish. Proceeding along similar lines, one can write the remaining $3$-forms in terms of the five independent ones $\{{\bf E}^{01\bar1}, {\bf E}^{02\bar2}, {\bf E}^{\bar001}, {\bf E}^{\bar00\bar1}, {\bf E}^{\bar01\bar1}\}$.

To compute the momenta, we need the derivatives of the Lagrangian with respect to the vielbeins. Let us do a sample computation to illustrate a few points 
\begin{multline}
\frac{\delta}{\delta {\bf E}^1}(-\det h)= 
-\frac{1}{8} \big({\bf E}^{0\bar01\bar{1}}+{\bf E}^{0\bar02\bar 2}\big){\bf E}^{0\bar0\bar1}\cr
+\frac{1}{4}\left[
-(\mathfrak{e}^0\wedge {\bf E}^{2\bar 2})(\mathfrak{e}^0\wedge {\bf E}^{\bar1\bar2 2})
+(\mathfrak{e}^9\wedge {\bf E}^{2\bar 2})(\mathfrak{e}^9\wedge {\bf E}^{\bar1\bar2 2}) 
+(\mathfrak{e}^0\wedge {\bf E}^{2\bar 1})(\mathfrak{e}^0\wedge {\bf E}^{\bar 2\bar1 1})\right.\cr
\left. -(\mathfrak{e}^0\wedge {\bf E}^{21\bar 1})(\mathfrak{e}^0\wedge {\bf E}^{\bar2\bar1}) 
-(\mathfrak{e}^9\wedge {\bf E}^{2\bar 1})(\mathfrak{e}^9\wedge {\bf E}^{\bar 2\bar11})
+(\mathfrak{e}^9\wedge {\bf E}^{21\bar 1})(\mathfrak{e}^9\wedge {\bf E}^{\bar2\bar1}) )\right]\,.
\end{multline} 
Using the BPS equations, one can write
\be
\mathfrak{e}^9\wedge {\bf E}^{\bar{1}\bar{2}2} = -\mathfrak{e}^0\wedge {\bf E}^{\bar{1}\bar{2}2} \,.
\ee
Substituting this into the above expression and using $\mathfrak{e}^0$ with $\frac{1}{2}({\bf E}^{\bar0}+{\bf E}^0)$, we get
\begin{align}
\frac{\delta}{\delta {\bf E}^1}(-\det h) 
&=\left[ -\frac{1}{8} \big({\bf E}^{0\bar0}\wedge({\bf E}^{1\bar{1}}+{\bf E}^{2\bar 2})\big){\bf E}^{0\bar0\bar1}+\frac{1}{8}\left[-{\bf E}^{02\bar2}({\bf E}^{\bar0\bar1\bar22})+{\bf E}^{02\bar1}({\bf E}^{\bar0\bar2\bar11})-{\bf E}^{0\bar2\bar1}({\bf E}^{\bar021\bar1})\right]\right] \,.
\end{align}
Now, observe that the last term in the above equation is zero since ${\bf E}^0$, ${\bf E}^{\bar1}$ and ${\bf E}^{\bar2}$ are linearly dependent. Using the relations between the various $3$-forms and $4$-forms and the expression for the determinant with the BPS equations are imposed,
\be
\sqrt{-\det\, h} = \mathfrak{e}^{09}\wedge(\mathfrak{e}^{13}+\mathfrak{e}^{24})= \frac{i}{4}\, ({\bf E}^{0\bar01\bar1}+{\bf E}^{0\bar02\bar2}) \,,
\ee
one can check that the above derivative can be re-written as
\begin{align}
\frac{1}{\sqrt{-\det h}}\, \frac{\delta}{\delta {\bf E}^1}(-\det h)&=
\frac{i}{2}\big({\bf E}^{0\bar0\bar1}+{\bf E}^{\bar12\bar2} \big)\,. 
\end{align}
Similar expressions can also be derived for the other derivatives. The contribution to the momenta from the WZ part of the action can be easily computed since it is a simple wedge product of the vielbeins. Restricting attention to only the DBI part of the action, and using the definition of the momenta,
\be
P_{a} =  -\frac{N}{4\pi^2l^4}\frac{1}{\sqrt{-\det h}}\, \frac{\delta}{\delta {\bf E}^a}(-\det h) \,,
\ee
we obtain the following expressions for the momenta:
\begin{align}
P_{1} &= -\frac{iN}{8\pi^2l^4}{\bf E}^{\bar1}\wedge({\bf E}^{0\bar0}+{\bf E}^{2\bar2}) \cr
P_{2} &= -\frac{iN}{8\pi^2l^4}{\bf E}^{\bar2}\wedge({\bf E}^{0\bar0}+{\bf E}^{1\bar1}) \cr
P_{\bar{0}} &= (p_0 - p_9)= \frac{iN}{4\pi^2 l^4}  {\bf E}^0 \wedge ({\bf E}^{1\bar 1} + {\bf E}^{2 \bar 2})\cr
P_{0} &= (p_0 + p_9)= \frac{iN}{4\pi^2 l^4}\left[-{\bf E}^{\bar0}\wedge({\bf E}^{1\bar1}+{\bf E}^{2\bar2})+{\bf E}^{0}\wedge\left(\left\vert\frac{a_0}{a_2}\right\vert^2{\bf E}^{1\bar1} +\left\vert\frac{a_0}{a_1}\right\vert^2{\bf E}^{2\bar2}\right) \right] 
\end{align}

These can be very simply related to the derivatives of the Lagrangian with respect to the real one-forms $\mathfrak{e}^a$ by taking linear combinations of the complex momenta. Combining this with the contribution from the WZ piece, we obtain the real momenta written out in the text. 

\section{Hyperbolic complex projective space $\widetilde{\mathbb{CP}}^m$}

The hyperbolic versions of the complex projective space can be defined as the set of rays in $\mathbb{C}^{1,N}$. One can check that this space is a K\"ahler manifold. We start with the coordinates $\{ \Phi_0, \Phi_1, \cdots, \Phi_m\}$ on $\mathbb{C}^{1,m}$ and define the new variables
\begin{eqnarray}
\xi = \Phi_0 \cdots \Phi_m\qquad b_k = \frac{\Phi_k}{\Phi_0}\, \quad \text{with} \quad k = 1, \cdots, m\,.
\end{eqnarray}
and the inverse transformations
\begin{equation}
\Phi_0^{m+1} = \frac{\xi}{b_1 \cdots b_m}\qquad \Phi_k^{m+1} = \frac{\xi}{b_1 \cdots b_m} \, b_k^{m+1} ~~ {\rm for} ~~ k = 1, \cdots, m.
\end{equation}
These imply the differential conditions
\begin{eqnarray}
\frac{d\Phi_0}{\Phi_0} &=& \frac{1}{m+1} \left[ \frac{d\xi}{\xi}-\sum_{k=1}^m \frac{db_k}{b_k} \right] \cr
\frac{d\Phi_k}{\Phi_k} &=& \frac{d\Phi_0}{\Phi_0} + \frac{db_k}{b_k} ~~~ \forall ~~~ k \in \{1, \cdots , m\}.
\end{eqnarray}
We will now impose the conditions
\begin{equation}
|\Phi_0|^2-|\Phi_1|^2- \cdots - |\Phi_m|^2 = 1\quad\text{and}\quad (\Phi_0 \Phi_1 \cdots \Phi_m) \in {\mathbb R}.
\end{equation}
to obtain the complex hyperbolic space ${\widetilde{\mathbb{CP}}}^m$. Using these conditions we can write $\xi$ in terms of $b_k$ as
\begin{equation}
\xi = \frac{|b_1 \cdots b_m|}{(1-\sum_{p=1}^m |b_p|^2)^{\frac{m+1}{2}}}.
\end{equation}
Using this we can write 
\begin{eqnarray}
\frac{d\xi}{\xi}  = \frac{1}{2} \left( 1+ \frac{m+1}{1-\sum_{p=1}^m |b_p|^2} |b_k|^2 \right) \left( \frac{d\bar b_k}{\bar b_k} + \frac{db_k}{b_k} \right).
\end{eqnarray}
Now, the K\"ahler form on ${\widetilde{\mathbb{CP}}^m}$ is inherited from that on $\mathbb{C}^{1,m}$:  
\begin{eqnarray}
\omega_{\widetilde{\mathbb{CP}}^m} = -i \, d\bar\Phi_a \, \wedge d\Phi_b \, \eta^{ab}
\end{eqnarray}
where $\eta^{ab} = {\rm diag} \{-1, +1, \cdots, +1\}$. Using the differential conditions obtained earlier, this works out to be
\begin{equation}\label{omegacptilde}
\omega_{\widetilde{\mathbb{CP}}^m} = -i  \left[ \delta_{mn} + \frac{ b_m {\bar { b}}_n}{1-\sum_{p=1}^m | b_p|^2} \right] \frac{d{\bar { b}}_m \wedge d b_n}{1-\sum_{p=1}^m | b_p|^2}
\end{equation}
which can be seen to be the K\"ahler form generated from the K\"ahler potential 
\be
K = -\ln (1-\sum_{p=1}^m |b_p|^2)\,,
\ee 
with $\omega_{\widetilde{\mathbb{CP}}^m} = -i \, \partial_{\bar m} \partial_n K \, d {\bar {b}}_m \wedge d b_n$.

\subsection{Geometric quantization of $\widetilde{\mathbb{CP}}^m$}\label{quantumcptilde}

We will be very brief in this section and refer the reader to \cite{woodhouse} for a general discussion of geometric quantization, in particular, holomorphic quantization of K\"ahler manifolds. 

We start directly with the symplectic $2$-form of interest on $\widetilde{\mathbb{CP}}^m$, which is given by 
\begin{equation}
\omega = -i \, N \frac{1}{(1-|b|^2)}\left(\delta_{ij} + \frac{b_i\, \bar b_j}{1-|b|^2}\right) d\bar b_i \wedge db_j \,,
\end{equation}
where $|b|^2 = \sum_{p=1}^{m} |b_p|^2$. This is $N$ times the symplectic form in \eqref{omegacptilde}. We choose to work with holomorphic polarization
\begin{equation}
D_{\bar b_i} \phi(b_i, \bar b_j) = 0
\end{equation}
where $\phi(b_i, \bar b_j)$ are the wave-functions. Then the adapted K\"ahler gauge potential is
\begin{equation}
\theta = -i \, N \frac{\bar b_i}{1-|b|^2} db_i
\end{equation}
and the K\"ahler potential is $K = -N \ln (1-|b|^2)$ so that $\theta = -i \partial_b K \, db$. Then the covariant derivative ($D_j = \partial_j  - i \, \theta_j$ with $\hbar$ set to unity)  $D_{\bar b}$ is simply $\partial_{\bar b}$ and 
\begin{equation}
D_{b_i} = \partial_{b_i} - \frac{N \, \bar b_i}{1-|b|^2} \,.
\end{equation}
The K\"ahler form $\omega$ is treated as an anti-symmetric matrix; it is non-degenerate and the inverse matrix, which is essential to define the Poisson brackets of functions on classical phase space, is given by
\begin{equation}
\omega^{\bar b_i b_j} = \frac{i}{N} (1-|b|^2)(\delta_{ij}-\bar b_i b_j) \,.
\end{equation}
The prescription for geometric quantization \cite{woodhouse} is to map functions on phase space to operators, with the map
\begin{equation}
f \rightarrow i \, \partial_i f \, \omega^{ij} \, D_j + f
\end{equation}
These are thought of as acting on the states in the Hilbert space (or wavefunctions) which satisfy the polarization condition $D_{\bar b}\phi = 0$. Substituting the explicit expression for the inverse of the K\"ahler form, the map from functions to operators takes the form
\begin{align}\label{fnopmap}
f &\rightarrow \frac{1}{N}\sum_{i,j} \partial_{\bar b_i}f (1-|b|^2)(\delta_{ij} - \bar b_i b_j)\left(\frac{\partial}{\partial b_j} - \frac{N\bar b_j}{1-|b|^2} \right) \cr
&\rightarrow \frac{1}{N} (1-|b|^2)\left(\sum_{i}\partial_{\bar b_i}f \frac{\partial}{\partial b_i} - \sum_{i,j} \bar b_i\partial_{\bar b_i}f \, b_j \frac{\partial}{\partial b_j} - N \sum_{i}\bar b_i \partial_{\bar b_i}f \right)
\end{align}
In computing the partition function for the 1/2-BPS dual-giants, it is necessary to find the differential operator representation of the angular momentum $J_1$. In terms of the $\zeta$-variables defined in \eqref{flatomega} describing the configuration space as $\mathbb{C}^N$, the classical expression for the conserved charge $J_1$ is given by
\be
J_1 = \sum_{k=1}^{N} k |\zeta_k|^2 \,.
\ee
Re-expressing this in terms of the $b$-variables using equations \eqref{cntocpn} and \eqref{fmap}, we find that 
\begin{equation}\label{Jdefb}
J_1(b, \bar b) = N \sum_{k=1}^{N}\frac{k\, |b_k|^2}{1-|b|^2} \,.
\end{equation}
The reason for the weighted sum in the expression for $J_1$ is similar to the one given in \cite{bglm}; the coordinates of $\widetilde{\mathbb{CP}}^N$ can be thought of as symmetric combinations of the zeroes of the defining polynomial of degree $N$. So, assigning unit degree to the roots of the polynomial, we see that the coordinates $b_k$ have charge $k$. We will verify this explicitly using the map that takes the function $J_1$ on phase space to an operator in the quantum theory. Substituting \eqref{Jdefb} into \eqref{fnopmap}, we get the operator 
\begin{align}
J_1&\rightarrow \sum_{k=1}^{N} k\, b_k\frac{\partial}{\partial b_k} \,.
\end{align}
We now turn to discuss the wavefunctions and their inner-product. Since the polarization is given by $D_{\bar{b}}\phi=0$, the wavefunctions are holomorphic in $b_i$.  A convenient basis is given in terms of monomials 
\be
\phi= \prod_{i}b_i^{p_i}\,.
\ee
Following the notations of \cite{bglm}, the function $W$, which is relevant for defining the inner product is 
\be
W = e^{-K} = (1-\vert b\vert^2)^N \,.
\ee
For instance, for the simple case of the poincare disc, or $\widetilde{\mathbb{CP}}^1$, the norm of such a basis element is therefore given by 
\begin{align}
\langle \phi_p, \phi_p \rangle &= (N-1)\int_{\widetilde{\mathbb{CP}}^1} \omega_{\widetilde{\mathbb{CP}}^1} \, \bar \phi_p\, \phi_p\, W\cr
& =(N-1)\int_{0}^{1} dr\, r^{2p+1} (1-r^2)^{N-2} \,.
\end{align}
The integral is finite for any integers $p \ge 0$ and $N \ge 2$. The extra factor of $N-1$ is put to ensure that one has constant wave-functions normalizable also for $N=1$ case \cite{berezin}.

The partition function for $\widetilde{\mathbb{CP}}^N$, given these wavefunctions and given the operator expression for the angular momentum, is given by
\be
Z = {\rm Tr}_{\cal H}\ e^{-\beta J_1} = \prod_{k=1}^{N}\frac{1}{1-q^k} \quad \text{with} \quad q = e^{-\beta} \,.
\ee
Note that one could as well have quantized $\mathbb{C}^N$ and gotten the same answer. 

\end{appendix}

\end{document}